\documentclass{article}

\usepackage{arxiv}

\usepackage[utf8]{inputenc} 
\usepackage[T1]{fontenc}    
\usepackage{hyperref}       
\usepackage{url}            
\usepackage{booktabs}       
\usepackage{amsfonts}       
\usepackage{nicefrac}       
\usepackage{microtype}      
\usepackage{lipsum}
\usepackage{bm}
\usepackage{graphicx}
\usepackage{multirow}
\usepackage{color,soul}
\usepackage{ulem,xpatch}
\graphicspath{ {./images/} }
\usepackage{float}
\usepackage{setspace}   
\usepackage{xcolor}
\usepackage{amsmath}
\usepackage{amssymb}
\usepackage{blindtext}
\usepackage{lineno}

\usepackage{newtxmath}

\usepackage[round]{natbib}   

\title{Experimental and numerical study of the effect of polymer flooding on sand production in poorly consolidated porous media}

\author{ \href{https://orcid.org/0000-0001-8885-9511}{\includegraphics[scale=0.06]{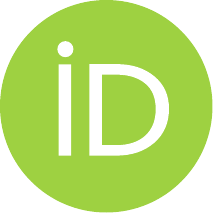}\hspace{1mm}Daniyar~Kazidenov }\\
	Department of Mathematics, \\
    School of Sciences and Humanities,\\
	Nazarbayev University\\
	Astana, Kazakhstan \\
	\texttt{daniyar.kazidenov@nu.edu.kz} \\
	\And
	\href{https://orcid.org/0000-0003-2277-7390}{\includegraphics[scale=0.06]{orcid.pdf}\hspace{1mm}Sagyn~Omirbekov}\thanks{Corresponding author}  \\
	Department of Mathematics, \\
    School of Sciences and Humanities,\\
	Nazarbayev University\\
	Astana, Kazakhstan \\
	\texttt{sagyn.omirbekov@nu.edu.kz} \\
  	\And
	\href{https://orcid.org/0000-0002-5242-0440}{\includegraphics[scale=0.06]{orcid.pdf}
    \hspace{1mm}Meruyet~Zhanabayeva} \\
	Petroleum Engineering,\\ 
    School of Mining and Geosciences,\\
	Nazarbayev University\\
	Astana, Kazakhstan \\
	\texttt{meruyet.bazhanova@nu.edu.kz} \\
 	\And
	\href{https://orcid.org/0000-0002-3958-8871}{\includegraphics[scale=0.06]{orcid.pdf}
    \hspace{1mm}Yerlan~Amanbek} \\
	Department of Mathematics,\\
    School of Sciences and Humanities, \\
	Nazarbayev University\\
	Astana, Kazakhstan \\
	\texttt{yerlan.amanbek@nu.edu.kz} \\
}

\renewcommand{\shorttitle}{\textit{arXiv} Template}

\hypersetup{
pdftitle={A template for the arxiv style},
pdfsubject={q-bio.NC, q-bio.QM},
pdfauthor={David S.~Hippocampus, Elias D.~Striatum},
pdfkeywords={First keyword, Second keyword, More},
}
\begin{document}
\maketitle
\begin{abstract}
	
    Polymer flooding is crucial in hydrocarbon production, increasing oil recovery by improving the water-oil mobility ratio. However, the high viscosity of displacing fluid may cause problems with sand production on poorly consolidated reservoirs.  
     This work investigates the effect of polymer injection on the sand production phenomenon using the experimental study and numerical model at a laboratory scale.

    The experiment uses an artificially made sandstone based on the characteristics of the oil field in Kazakhstan. Polymer solution based on Xanthan gum is injected into the core to study the impact of polymer flooding on sand production. The rheology of the polymer solution is also examined using a rotational rheometer, and the power-law model fits outcomes. The fitting parameters are used for the numerical model as an input. We observe no sand production during the brine injection at various flow rate ranges. However, the sanding is noticed when the polymer solution is injected. More than 50\% of cumulatively produced sand is obtained after one pore volume of polymer sand is injected.
    
     In the numerical part of the study, we present a coupling model of the discrete element method (DEM) with computational fluid dynamics (CFD) to describe the polymer flow in a granular porous medium. The numerical model is performed considering particle size distribution, porosity, and cementation behavior of the sample associated with Kazakhstan reservoir sandstone. In the solid phase, the modified cohesive contact model characterizes the bonding mechanism between sand particles. The fluid phase is modeled as a non-Newtonian fluid using a power-law model. The forces acting on a particle by the fluid are calculated considering the rheology of non-Newtonian fluid.
     We verify the numerical model with the laboratory experiment by comparing the dimensionless cumulative mass of produced particles. The numerical model observes non-uniform bond breakage when only a confining stress is applied. On the other hand, the injection of the polymer into the sample leads to a relatively gradual decrease in bonds. The significant difference in the pressure of the fluid results in its higher velocity, which causes intensive sand production at the beginning of the simulation. During the transient phase in sand production, the fluid's viscosity is lower at the outlet region, where the unbonded particles significantly predominate. The ratio of medium-sized produced particles is greater than the initial ratio of those before injection and makes the most significant contribution to the total mass of sand production.   \\

\end{abstract}


\section{Introduction}
Weakly consolidated sandstone reservoirs occur in a significant fraction of the world's oil and gas reserves, making them sensitive to sand production. The destruction of the material is an essential step in the grinding process. Drilling operations, shut-in and start-up cycling, operating conditions, reservoir pressure reduction, and water weakening can all contribute to the destruction of sandstone near perforations and wells. Separation of the sand particles is also facilitated by the high-pressure gradient caused by the fluid flow. In addition, the fluid flow is responsible for transporting and forming loose sand particles or loose sand lumps into the wellbore. Therefore, the sand production negatively affects the well production by damaging the equipment due to the erosion, reducing the flow diameter and flow rates, and clogging of wellbore and surface facilities \citep{rahmati2013review}. Depending on the behavior of the sand production rate, the process can be in three different regimes: transient, continuous and catastrophic sand production \citep{veeken1991sand}. While during the transient sand production the total mass of produced sand decreases with time, during the continues regime the mass of produced sand has constant pattern and does not change over the time. The catastrophic sand production causes the failure of the wellbore by sudden blocking the lines.

Oil production stages such as primary and secondary methods (i.e., spontaneous imbibition and water injection, respectively) can affect sand production during oil recovery.  
Terzaghi conducted the first scientific studies on the sanding problem in 1936, and he was the first to notice a sand arch near a bottom hatch in a box filled with sand \citep{terzaghi1936stress}. Terzaghi's experiment was enhanced by \cite{hall1970stability}, who discovered a link between the added fluid flow and the creation of sand arches. Later experiments with various hydrocarbon production parameters indicated that sand production is influenced by stress anisotropy, stress level, saturation, injection fluid, and rock formation material. Many researchers have also experimented with a variety of fluids, including water \citep{wang2001borehole, li2019sand}; diesel fluid \citep{fattahpour2012experimental}; brine \citep{kooijman1996horizontal} and paraffin oil \citep{papamichos2001volumetric}. 

Moreover, it is essential to note that each field is unique, and the sand production can vary depending on the specific reservoir characteristics, such as the geomechanical properties of the formation  \cite{kozhagulova2021integrated}, the location of weak zones \cite{shabdirova2020experimental}, and reservoir fluid type \cite{khamitov2022numerical}. Precisely, \cite{kozhagulova2021integrated} experimentally investigated the sand production behavior in poorly consolidated formations. The artificially synthetic sample that replicates the reservoir sandstone from Kazakhstan, was created with sodium silicate cementing agent. The patterns of sand production that were observed in the laboratory were comparable to the real behavior of sand production in the local fields, where a critical flow rate initiated sand production and a small burst of sand occurred with the subsequent increase in flow rate. \cite{shabdirova2020experimental} performed experimental and numerical analysis to understand the change in the permeability of the plastic zone surrounding the wellbore, which in turn provided valuable insight into the mechanical and filtration characteristics of the rock. The findings demonstrated that the average permeability through the sample dropped as the flow rate increased. \cite{khamitov2022numerical} developed a numerical modeling to investigate the effect of various reservoir fluids on sand production. It was observed that heavy oil produced more sand than light oil, due to its greater transport capacity and its creation of a more uniform particle velocity trajectory pattern. All these previous studies on Kazakhstani oil fields considered the effect of only Newtonian fluids on sand production.

Polymer flooding is a method of oil production that uses synthetic or bio polymers to increase oil production by improving the mobility ratio between displacing fluid and oil \citep{mandal2015chemical}. Hence, the oil displacement by aqueous solution occurs due to increased water viscosity by adding polymers. The polymer flooding can cause significant challenges on sand production control, specifically, when oil is recovered from poorly consolidated reservoirs. Nevertheless, the effect of polymer injection (i.e., enhanced oil recovery, EOR method) on sand production is still unclear since only a few research have been conducted in this field. For instance, \cite{guo2019lessons} pointed out a severe sand production problem during the alkali/surfactant/polymer (ASP) flooding in the poorly consolidated Shengli Oil Field, where the ASP flooding produced ten times more sand than common water flooding. \cite{li2005comparison} also explored the sanding phenomenon in the field test when they injected foam. The foam was generated by co-injecting Nitrogen and surfactant-polymer solution. They also noticed the production of sand in central wells where after adding perforation and sanding control, water cut declined, and oil production rose significantly. Some authors mentioned that polymer flooding could minimize sand production by decreasing the water hammering effect due to higher viscosity of polymers \citep{bautista2017state}. Therefore, the sweeping efficiency is improved which stabilizes the pressure in the formation. 

Numerous approaches have been also developed to understand the mechanisms of sand production including analytical studies, laboratory experiments and numerical models. While conventional laboratory testing and analytical models can only predict the beginning of sand production, more advanced laboratory experiments and numerical models might predict volumetric sand production rate \citep{rahmati2013review}. For instance, \cite{willson2002new} proposed an analytical model to predict the sand production rate, which works only in continues regime. Derived empirical correlations between  loading factor, Reynold’s number and the sand production rate using laboratory experimental data showed a good match with the results of field data. \cite{papamichos2001volumetric} experimentally developed a mechanical-erosion model to predict volumetric sand production rate in weak and compactive sandstones. Externally imposed stress caused decohesioning and plasticification of a zone around the cavity where erosion occurred due to fluid flow. The experimental results showed that the sand production appears to remain constant over time under constant external stress and flow rate. Coupling of computational fluid dynamics (CFD) and discrete element method (DEM) \citep{tsuji1993discrete} is one of the common numerical approaches used to model sand production phenomena, in which the sand grains are considered as discrete particles, while the fluids are considered as a continuous phase. \cite{zhou2011numerical} investigated sand erosion in a weakly cemented sandstone using a CFD-DEM coupling. An increase of the axial compaction accelerated the sand erosion,  and an increase of radial confining stress resulted in continuous sand production. Moreover, the fluid- particle interaction force were the primary cause of sand erosion. CFD-DEM is also helpful to understand the interaction mechanism of sand particles with fluid that is affected by many factors such as different fluid flow rates \citep{ma2021cfd}, particle size distribution \citep{zhao2013coupled}, cementation behaviour \citep{fang2022cfd} and wellbore geometry \citep{zhang2019proppant}. Moreover, CFD-DEM can be employed to simulate the fluid flow and motion of sand particles in complex geometries including gravel packs \citep{wang2022microscopic} and sand screen \citep{ismail2021cfd}.
 
The JKR model, which was initially proposed by \cite{johnson1971surface}, is a commonly used technique in CFD-DEM simulations to describe the bonding mechanism of adhesively contacted sand grains \citep{zhou2022calibration}. 
The force involved in the adhesion of the particles is expressed by the surface energy density and the contact area of those particles. JKR-based DEM modeling is a popular instrument to investigate various applications such as compaction of powders \citep{garner2018study}, micromanipulation \citep{fang2006dynamic}, behaviour of particles in microfluidics \citep{shahzad2018aggregation} and analysis of soil mechanics \citep{qiu2022calibration}.
\cite{rakhimzhanova2019numerical} proposed a modified version of the JKR model, which was applied to investigate the  bonding behavior of poorly consolidated cemented sandstone from a Kazakhstan reservoir. In contrast to the original JKR model where the bonds break as they pass the maximum applied force, the modified JKR model defines the bond breakage at a maximum force.  The bonded particles fracture in brittle manner due to the absence of tensile bonds. Furthermore, the broken particles are unable to create new bonds and modeled by the Hertz contact model \citep{hertz1882ueber}. The modified version of JKR model is successfully validated and used to investigate the the bond breakage in sand production \citep{rakhimzhanova2022numerical,khamitov2021coupled}, triaxial compression test {\citep{rakhimzhanova2019numerical, kazidenov2022coarseconf} and cone penetration test \citep{rakhimzhanova2021numerical}.

Recently, \cite{li2023simulation} presented another application of the CFD-DEM for non-Newtonian fluids in which they predicted fluid-induced fractures in granular media due to polymer injection by accounting water quality and polymer rheology. The polymer injection model assumes a homogeneous porous medium in a granular system containing spherical particles. The authors use a power law rheological model to simulate polymer flow behavior. \citep{frungieri2022cfd} employed CFD-DEM to investigate the breakup dynamics of solid fillers in polymer dispersing medium. While the non-Newtonian fluid with shear thinning behaviour is characterized by a power-law relationship, the solid fillers are described as micron-sized spherical particles interacting by van der Waals forces. However, a complete understanding of sand production in poorly consolidated porous media by polymer flooding is still unclear due to the complex structure of porous media.  

To our knowledge, the experimental and numerical assessment of sand production by polymer flooding has received less attention because of the complex (non-Newtonian) behavior of polymer solution and coupling with poorly consolidated porous media. Specifically, it is relevant for some Kazakhstani oil fields where porous media is weakly consolidated.

In this study, we focus on assessing sand production by polymer flooding in poorly consolidated formations of the Ustyurt-Buzachi sedimentary basin located in western Kazakhstan, see Figure \ref{fig:map}. The studies on sand production in the areas mentioned earlier by tertiary oil recovery methods are still unclear since most of the oil recovery methods fall under primary or secondary production techniques.  Consequently, the main objective of this investigation is experimentally and numerically show the impact of sand production by polymer flooding, specifically by a non-Newtonian fluid.  The ultimate goal is to assess and prevent the risk of sand production by polymer flooding for future projects.

\begin{figure}[H]
\begin{centering}
\includegraphics[width=0.5\columnwidth]{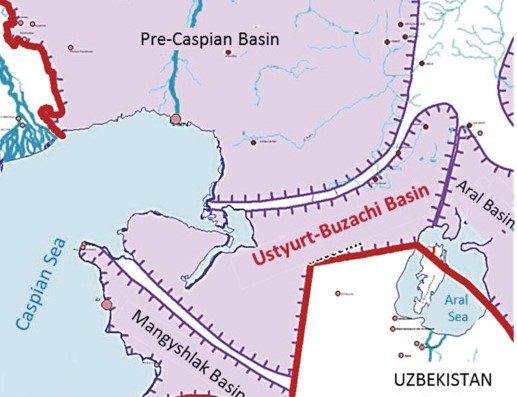}
\par\end{centering}
\caption{Ustyurt-Buzachi sedimentary basin. Adapted from \citep{iskaziyev2013complex}.\label{fig:map}}
\end{figure}

This paper presents the experimental and numerical study of polymer flooding in poorly consolidated porous media at a laboratory scale. The laboratory experiment is conducted using an artificial sandstone based on the particle size distribution of a Kazakhstani oil field to investigate the effect of polymer flooding on sand production rate. The viscosity of the polymer solution is measured using a rotational rheometer and is fitted by the power law model. In the numerical simulation, while the modified JKR-based DEM model characterizes the cementation behavior of particles, the power-law-based CFD model describes the polymer flow as a non-Newtonian fluid. The particle-fluid interaction forces are computed considering the rheology of non-Newtonian fluid. The numerical model is verified by the results obtained from the laboratory experiment. The fluid velocity and viscosity, cumulative sand production rate, particle bonding behavior, and particle size distribution of produced particles are examined in the effect of polymer injection on sand production.  

This work is organized as follows: in Section \ref{sec:exp}, we present experimental part of the work such as selecting the materials, experimental setup and procedure. The numerical part including governing equations and numerical simulation setup are presented in Section \ref{sec:numerical}. In Section \ref{sec:results}, we demonstrate and discuss the results obtained from the experiment and numerical simulation. Finally, the main findings are summarized in Section \ref{sec:conclusion}

\section{Experimental study}
\label{sec:exp}
We conducted experiments with poorly consolidated porous media using a core flood system to study the impact of polymer injection on sand production. Core samples were prepared at laboratory conditions following the field's Particle Size Distribution (PSD) and mechanical properties. This section presents the materials and experimental methods used to conduct the experiments. All materials were prepared according to the characteristics of the North Buzachi oil field.

\subsection{Materials}



\subsubsection{Porous media}
The artificial core samples were prepared using quartz sand provided by KazQuartz company. The sand is sieved and followed to obtain the same PSD as the reservoir sand, as presented by Rakhimzhanova and coworkers (Rakhimzhanova, 2021; Shabdirova, et al., 2016). The PSD of the reservoir sand is presented below in Figure \ref{fig:PSD}. 

\begin{figure}[H]
\begin{centering}
\includegraphics[width=0.6\columnwidth]{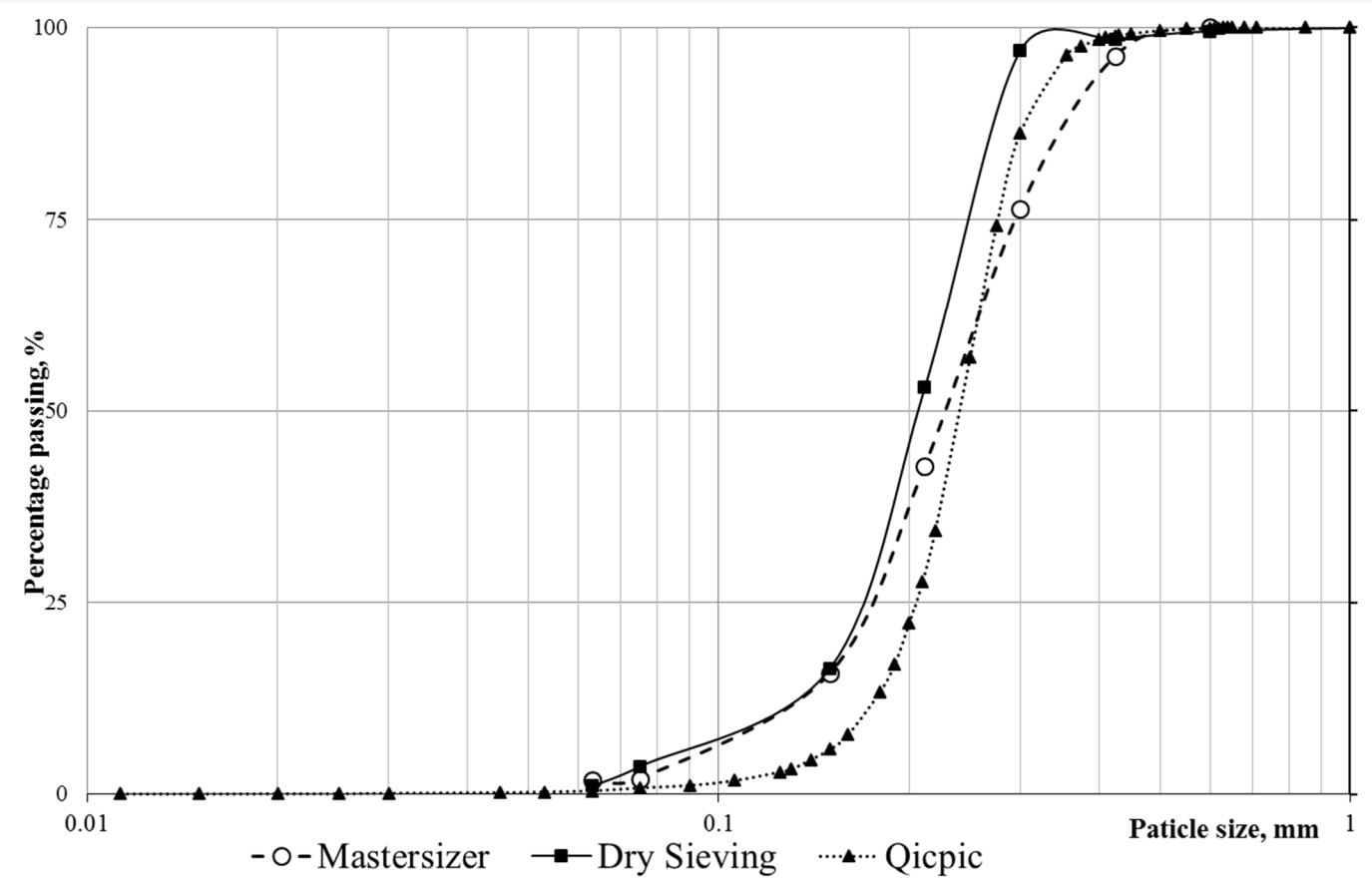}
\par\end{centering}
\caption{Particle Size Distribution of the reservoir examined through different techniques \citep{rakhimzhanova2019numerical}.\label{fig:PSD}}
\end{figure}

We used Portland cement, gypsum, and demineralized water to prepare the sandstones. An artificial sandstone with the oil field's mechanical properties was designed following Bisserik et al. (2021) method. The properties of the sandstone are tabulated in the following table. 

\begin{table}[H]
\fontsize{10}{12}\selectfont \caption{Property of prepared artificial sandstone.}
\begin{centering}
    \begin{tabular}{ p{3cm} p{2cm} p{2cm}  p{2cm} p{3cm}}
\toprule 
Core name & Sand (g) & Water (g) & Gypsum (g) & Portland cement (g) \\
\midrule
 & 957.0 & 156.2 & 20.1 & 46.9 \\
 \bottomrule
\end{tabular}
  \label{tab:core_parameters}
\par\end{centering}
\end{table}

To prepare sandstone with a certain consolidation the materials presented in Table \ref{tab:core_parameters} were very mixed and molded into a 120 mm long plastic cylinder with an internal diameter of 38.1 mm. The sandstones were dried under humidity of 80\% and at 60°C for 72 hours. The prepared sandstone presented in Figure \ref{fig:core} are then used for core flooding experiments.

\begin{figure}[H]
\begin{centering}
\includegraphics[width=0.6\columnwidth]{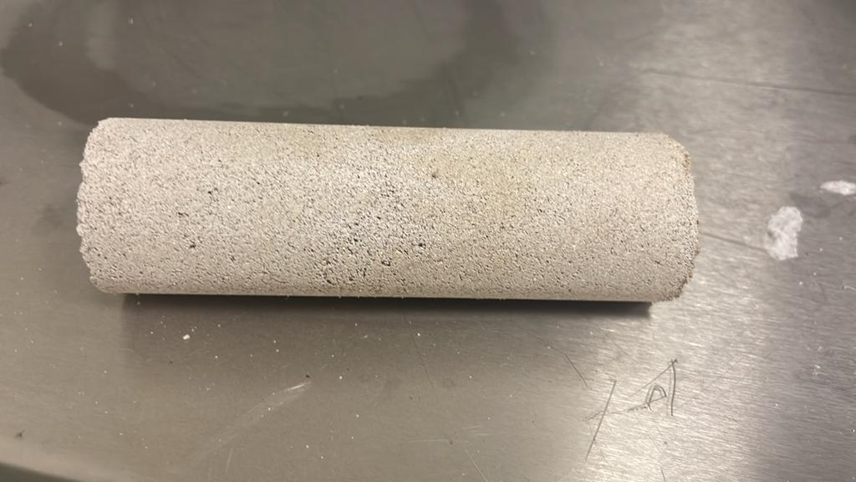}
\par\end{centering}
\caption{Artificially made sandstone core.\label{fig:core}}
\end{figure}

The porosity of sandstone is measured by a Helium porosimeter (Vinci technology). We saturated the cores with formation brine under pressure of 1500 psi for 48 hours to establish initial wetting conditions using a Manual Saturator (Vinci technology). The porosity was also rechecked by the dry/wet weight method. The measured properties of the artificial cores are presented in Table \ref{tab:pm_parameters}.

\begin{table}[H]
\fontsize{10}{12}\selectfont \caption{Property of prepared artificial sandstone.}
\begin{centering}
    \begin{tabular}{ p{3cm} p{3cm} p{2cm}  p{3cm} }
\toprule 
Name & Permeability, $K$ ($\mathrm{m^2}$) & Porosity, $\phi$ & Pore Volume ($ \mathrm {mL}$) \\
\midrule
 Core-7 & $74 (\pm{20}) \times\ 10^{-12}$ & $38.45\pm{1}$ & $45.55\pm{2}$ \\
 \bottomrule
\end{tabular}
  \label{tab:pm_parameters}
\par\end{centering}
\end{table}

\subsubsection{Formation brine}

The formation brine was prepared according to the chemical composition of water from the North Buzachi oil field. The total salinity of formation brine from the field is 93000 ppm. The formation brine contains sodium, calcium, magnesium, and chloride ions. The formation brine was prepared by adding salts in certain concentrations: $NaCl$ (63.24 g/L), $CaCl_2$ (12.34 g/L), and $MgCl_2 \cdot 6H_2O$ (37.52 g/L) to the distilled water and stirring at 750 RPM for 10 minutes. The density of formation brine is equal to 1.1166 g/L.

\subsubsection{Polymer solutions}

Xanthan Gum (XG) is employed as a viscous fluid to displace the initially saturated media by brine and investigate the sand production phenomenon. The XG powder is supplied by Sigma-Aldrich company. The polymer concentration was chosen to be 0.4 \% according to the outcomes of a previous study \citep{giese2002using} and Lenormand phase diagram to avoid any instabilities associated with capillary and viscous fingerings \citep{lenormand_touboul_zarcone_1988}. The viscosity is examined using a DHR-1 rheometer provided by TA Instruments following the previous study \citep{omirbekov2023experimental}. A required amount of fluid was loaded into the cone-and-plate geometry to fill the gap. We recorded the shear rate as a function of time corresponding to the given stress. The rheometer obtained each value when the shear stress change was less than a certain tolerance. Each shear rate was studied for 120 seconds from 0.1 to 100 1/s, and the number of experiments was triplicated. We plotted the shear viscosity ($\mu$}) versus shear rate ($\dot{\gamma}$) in Figure \ref{fig:XG}, where the XG polymer solution behaved as a shear-thinning, non-Newtonian fluid.

\begin{figure}[H]
\begin{centering}
\includegraphics[width=0.6\columnwidth]{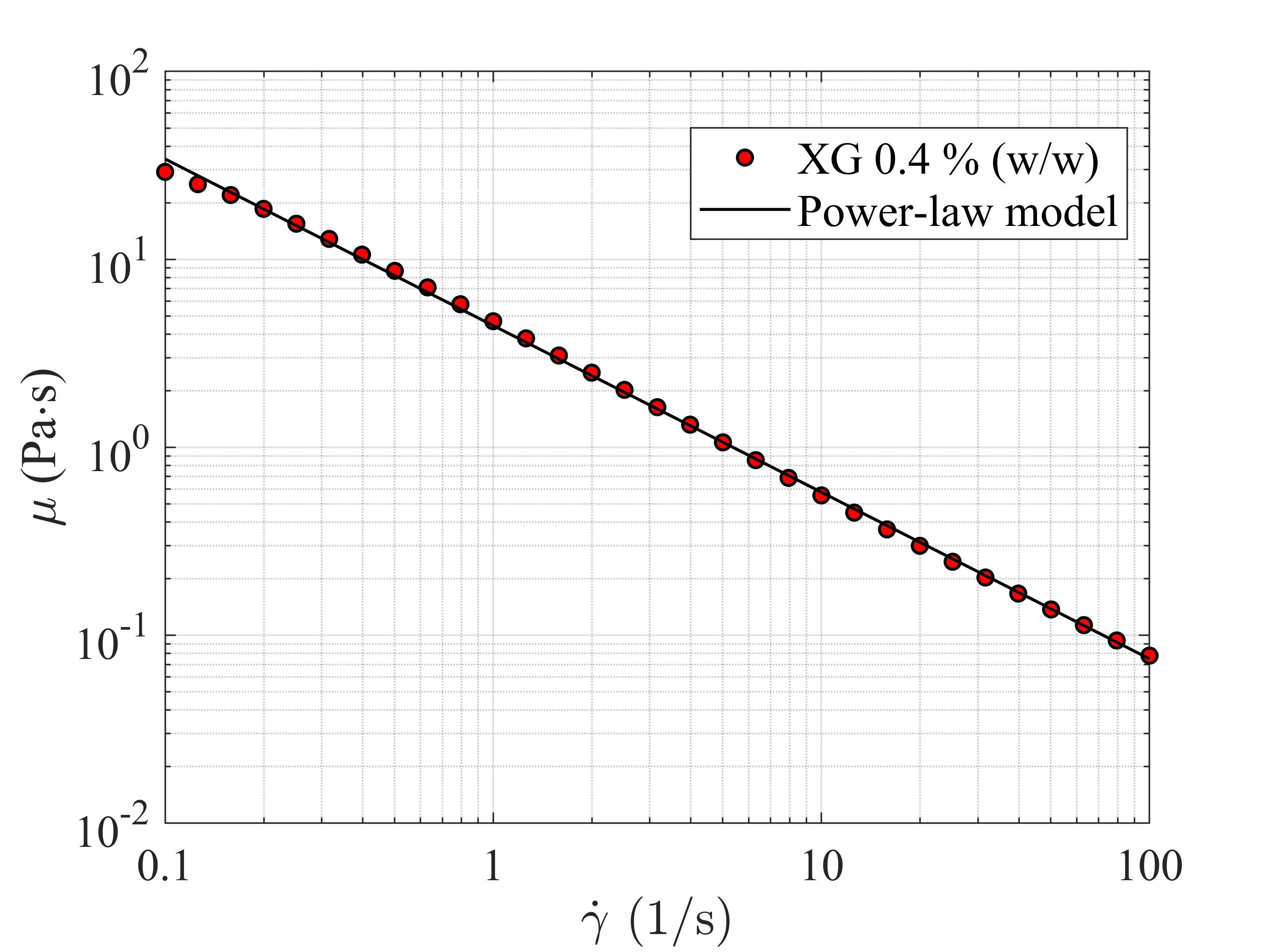}
\par\end{centering}
\caption{Shear viscosity versus shear rate of XG solution (0.4\% w/w) and fitting curve by Power-law model.\label{fig:XG}}
\end{figure}


The results are well fitted by Power-law model which is expressed as, 

\begin{equation}
\mu (\dot{\gamma}) = k\dot{\gamma}^{(n-1)} \label{eq:power-law}
\end{equation}
where $k$ (kg/m$\cdot$s) and $n$ (-) are the consistency and the flow indexes, respectively. We estimated the fitting parameters by nonlinear regression of the Matlab curve fitting toolbox.
The corresponding fitting parameters with the coefficient of determination is listed in Table \ref{tab:PL_param}. This model with parameters is used to model the polymer solution flow in a porous medium presented in the following sections.

\begin{table}[H]
\fontsize{10}{12}\selectfont \caption{Fitting parameters of the power-law model for XG solution (0.4\% w/w)}
\begin{centering}
    \begin{tabular}{ p{8cm} p{2cm}}
\toprule 
 Consistency index $k$,                               & $4.78$ \\
 Power-law index $n$                                  & $0.1547$ \\
 $R^2$                                                & $0.99$\\
\bottomrule
\end{tabular}
  \label{tab:PL_param}
\par\end{centering}
\end{table}

\subsection{Experimental setup}
\begin{figure}[H]
\begin{centering}
\includegraphics[width=0.6\columnwidth]{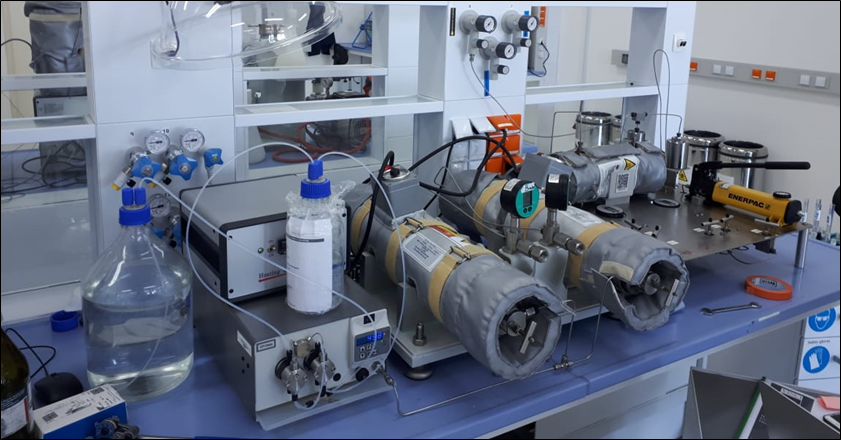}
\par\end{centering}
\caption{Experimental setup, ACA-700 Aging cell apparatus.\label{fig:aging_cell}}
\end{figure}

ACA-700 aging cell apparatus provided by VINCI technologies (see Figure \ref{fig:aging_cell}) is used to conduct core flooding experiments. It consists of a core holder, two accumulators (for injection fluid), a pump, pressure transmitters, and pressure regulators to set confining and back pressures. 

\subsection{Experimental procedure}

The formation brine was introduced horizontally into the core holder with a 3 mL/min flow rate to flush the core and measure the permeability. The permeability was calculated using Darcy's law, where we used values of the estimated pressure change along the core and the given flow rates.  
After, we injected the polymer solution into the core horizontally. The pressure difference of the fluid flow in the column was examined by gauging the pressure drop with pressure transmitters. The effluent was collected in graduated cylinders to assess sand production. Several graduated cylinders were used during the polymer injection. After every one PV of the polymer solution was injected, the effluent volume was collected in different graduated cylinders. As a result, the sand recovery was estimated using the effluent mass and volume difference. The radial confining pressure was set to 500 psi during all the experimental procedures conducted at 25°C in a temperature-controlled laboratory.

\section{Numerical simulation}
\label{sec:numerical}

\subsection{Model of the solid phase}

The DEM, initially introduced by \cite{cundall1979discrete}, is used to simulate the solid phase of the sand production model by solving Newton’s second law to describe the motion of individual particles and their trajectories. The particles in the systems accelerate due to inter-particle collisions, gravity, and fluid-particle interactions. The governing equations for the translation and rotation of particles are expressed as follows:


\begin{align}
m_i\frac{d \bm{v}_i}{dt} &= \bm{f}_{pf,i} +\sum_{j=1}^{k_c} \left( \bm{f}_{c,ij} + \bm{f}_{damp,ij} \right) +m_i \bm {g}\label{eq:1} \\
I_i\frac{d \bm{\omega}_i}{dt} &= \sum_{j=1}^{k_c} \bm{T}_{ij} \label{eq:2}
\end{align}

where $\bm{v}_i$ and $\bm{\omega}_i$ are the particle translational and angular velocities, $m_i$ and $I_i$ are the particle mass and inertia, $\bm{f}_{pf,i}$ is the particle-fluid interaction force,  $\bm{f}_{c,ij}$ and $\bm{f}_{damp,ij}$ are the contact and viscous damping forces between interacting particles, $m_i \bm g $ is the gravitational force, $\bm{T}_{ij}$ is the torque acting between particles $i$ and $j$ and $k_c$ is the number of interacting particles. The particle motion is influenced by the impact from fluid (in the presence of fluid), interaction with other particles and gravity. The particle-fluid interaction force  $\bm{f}_{pf,i}$ is the sum of all forces acting on a particle by fluid: 

\begin{equation}
\bm{f}_{pf, i} = \bm{f}_{d, i} + \bm{f}_{\nabla p, i} + \bm{f}_{\nabla \cdot \bm {\tau}, i} + \bm{f}_{Ar, i} \label{eq:2b}
\end{equation}

where $ \bm f_{d,i}$ is the drag force, $  \bm f_{\nabla p,i}$ is the pressure gradient force, $ \bm f_{\nabla \cdot \bm{\tau},i}$ is the viscous force and $ \bm f_{Ar, i}$ is the Archimedes force.


The particle-particle interaction forces are calculated using the linear spring-dashpot-slider model \citep{cundall1979discrete} which describes the contact force $\bm{f}_{c,ij}$:

\begin{equation}
\bm{f}_{c, ij}^{(n)} = - k_{n} \bm{\delta}_{n, ij} - \eta_n \bm{v}_{n, ij} \label{eq:3}
\end{equation}

\begin{equation}
\bm{f}_{c, ij}^{(t)} = - min \left(\mu \bm{f}_{c, ij}^{(n)}, k_{t} \bm{\delta}_{t, ij} + \eta_t \bm{v}_{t, ij} \right) \label{eq:4}
\end{equation}


where the superscripts $n$ and $t$ refer to normal and tangential, $k$ is the spring stiffness constant, $\bm{\delta_{ij}}$ is the overlap between the particles in contact, $\bm{v_{ij}}$ is their relative velocity, $\mu$ is the slider friction coefficient, and $\eta$ is the dashpot damping coefficient. 


The JKR contact model describes the adhesion and deformation behavior between cohesively bonded particles \citep{johnson1971surface}. It is represented in terms of contact force in normal direction:

\begin{equation}
f^{(n)}_{JKR} = \frac{4E^{*}a^{3}}{3R^{*}}-\sqrt{16\pi\gamma E^{*}a^{3}}\label{eq:12}
\end{equation}
where $E^{*} = \left(\displaystyle \frac{1 - \nu^{2}_{1}}{E_{1}} +\displaystyle \frac{1 - \nu^{2}_{2}}{E_{2}}\right)^{-1}$ is the effective Young's modulus, where  $E_{1}$, $E_{2}$ are the Young's modulus and $\nu_{1}$, $\nu_{2}$ are the Poisson's ratios of the particles, $\gamma$ is the surface energy density, $a$ is the radius of the contacting surface, and $R^{*} = \left(\displaystyle \frac{R_{i}R_{j}}{R_{i}+R_{j}}\right)$ is the effective radius, where  $R_{i}$ and $R_{j}$ are the particle radii that are in contact. 

The JKR model is based on the well-known Hertz \citep{hertz1882ueber} and Mindlin \citep{mindlin1949compliance} contact model and characterizes the adhesive bonding behavior of the particles. In equation \ref{eq:12}, while the first term represents the Hertz force, which is $k_{n} \bm{\delta}_n = \frac{4E^{*}a^{3}}{3R^{*}}$, and the second term is the adhesion force.The adhesive force acting in the contact area leads to deformation of the contact surface. Therefore, the contact area in the JKR model usually stretches more extensively than the Hertz and Mindlin model. This is because the JKR model considers the surface energy density and interfacial adhesion characteristics of the particle surfaces, which affect the contact area and deformation. Therefore, the JKR model is capable of simulating large deformations with full surface contact where the overlap exceeds the particle radius.

\subsection{CFD-DEM coupling}

\cite{zhu2007discrete} proposed the model A approach of the CFD-DEM coupling, in which the fluid invades only the porous regions ($\alpha_f $) of the material. The fluid phase is described using the locally averaged Navier-Stokes equations, where the forces acting on a particle by the fluid are shared between solid and fluid phases: 

\begin{equation}
 \left\{
\begin{array}{l}
\frac{\partial \alpha_f}{\partial t} + \nabla\cdot (\alpha_f \bm{u})=  0 \\ \vspace{0.01cm} \\
\frac{\partial (\rho_f \alpha_f \bm{u})}{\partial t} + \nabla \cdot \left( \rho_f \alpha_f \bm{u}\bm{u}\right)=  -\alpha_f \nabla p +\alpha_f \nabla \cdot \bm{\tau} + \rho_f \alpha_f \bm{g} + \bm{F}^{A}_{pf}  
 \label{eq:5} \\
 \end{array} 
  \right.
\end{equation}

where $\rho_f$ is the fluid density, $\alpha_f$ is the volume fraction occupied by the fluid, $\bm{u}$ is the velocity, $p$ is the fluid pressure. The stress tensor is given by
$\bm{\tau}=\mu\left( (\nabla \bm{u})+(\nabla \bm{u})^T\right)$, where $\mu$ is the fluid dynamic viscosity and $\bm{F}_{pf}^{A}=\dfrac{1}{\Delta V}\sum_{i=1}^{n} \left( \bm f_{d,i} +  \bm{f}^{''}_i \right)$  is the volumetric particle-fluid interaction force, where $ \bm{f}^{''}_i$ is the sum of forces other than drag, pressure gradient and viscous forces, $\Delta V$ is the volume of a fluid cell. 



The interaction force between a particle and fluid is an essential part of the CFD-DEM coupling, since it has a considerable impact on particle motion, which in turn can influence the fluid flow behavior. The pressure gradient force is given by $ \bm f_{\nabla p,i} = - \nabla p \cdot V_{p,i} $ and viscous force is  $ \bm f_{\nabla \cdot \bm{\tau},i} = - (\nabla \cdot \bm{\tau}) V_{p,i} $, where $V_{p,i} $ is the volume of a single particle. The particle drag force depends on the particle size and shape, fluid properties, relative velocity between particle and fluid,  and drag coefficient. The general form of the drag force is expressed as follows:
\begin{equation}
\bm{f}_{d,i} = \pi d^2_{p,i}\left (\displaystyle\frac{1}{8}  \rho_f |\bm{u_i}-\bm{v_i}| (\bm{u_i} -\bm{v_i}) \right)C_{d,i} \label{eq:8}
\end{equation}

where $d_{p, i}$ is the particle diameter, $\bm{u_i}$ and $\bm{v_i}$ are the fluid and particle velocities and $C_{d, i}$ is the drag coefficient.

The drag coefficient is a function of particle Reynolds number, that can be characterized by the fluid properties and its viscosity. For a creeping flow of a power-law fluid, the drag coefficient of a spherical particle is expressed taking into account the rheology of a non-Newtonian fluid \citep{dazhi1985drag}: 
\begin{equation}
C_{d, i} = \frac{24 \chi(n)}{{Re_{p, i}}} \label{eq:9}
\end{equation}
where $Re_{p, i}$ is a particle Reynolds number, $n$ is power-law index and $\chi (n)$ porosity correction factor \citep{renaud2004power}:
\begin{equation}
\chi (n) = 6^{(n-1)/2} \left(\displaystyle \frac{3}{n^2 + n + 1} \right)^{n+1} \label{eq:10}
\end{equation}

The particle Reynolds number is given as follows \citep{atapattu1995creeping}:

\begin{equation}
Re_{p, i} = \frac{\rho_f d_{p, i}^n |\bm{u_i}-\bm{v_i}|^{2-n}}{k} \label{eq:11}
\end{equation}

where $k$ is a the consistency index.




\subsection{Numerical setup of the simulation}

\begin{figure}[H]
\begin{centering}
\includegraphics[width=0.6\columnwidth]{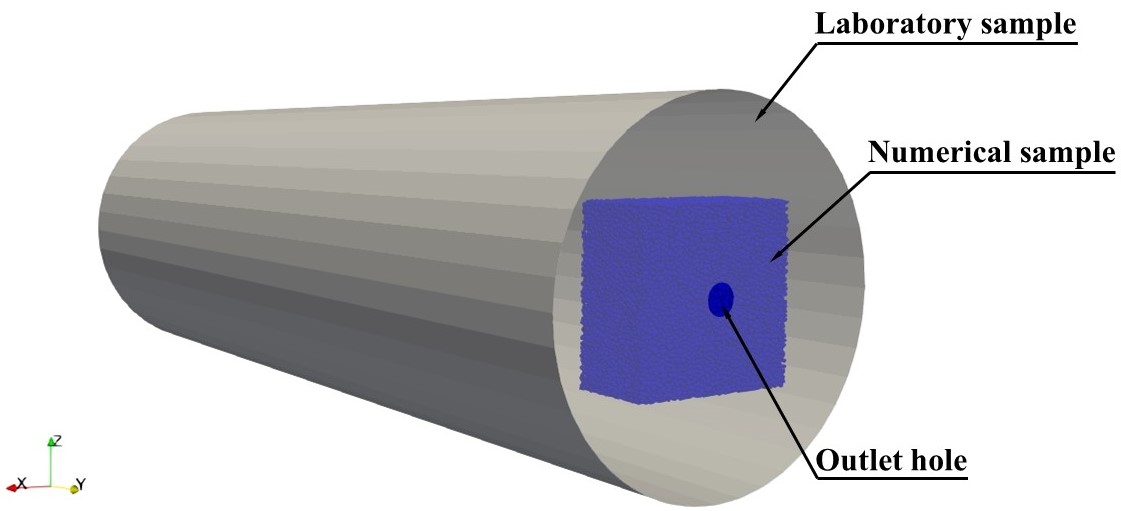}
\par\end{centering}
\caption{Geometry of the model compared to the sample in the experiment.\label{fig:sample}}
\end{figure}

In the numerical model, we reproduce the laboratory experiment (see Figure \ref{fig:core}), in which the sample is in 120 mm length and 38.1 mm diameter cylinder form experiencing the radial confining stress of 500 psi (Figure \ref{fig:sample}). The core used in the laboratory experiments contains approximately a billion small particles, which is costly for modeling as it requires a longer simulation runtime.Therefore, we use a small representative in the numerical model to reduce the computing power. Due to the software limitations, we create the numerical sample as a cuboid, with dimensions of 17 mm in length and height and 8.8 mm in width. The outlet hole has a diameter of 3.175 mm, as in the experiment. We assume that the stress applied on the experimental sample is equal to the stress exerted on the side planes of the numerical sample. To ensure consistency between the numerical simulation and laboratory experiment, we locate the numerical sample so that its outlet hole corresponds to the experimental one regarding both position and dimensions. 

The numerical simulation consists of separate DEM and CFD systems. The DEM system domain is constructed as a cube with six planes on all sides. These planes are servo wall type and contain and compress the particles in the domain. Initially, the particles are generated in the domain with specific predefined material parameters, dimensions, and geometry. At this stage, the Hertz contact model \citep{hertz1882ueber} eliminates any cohesion effect between particles. All particles in the system have a spherical shape with different PSD. Due to the small size and complexity of the particles, the simulation with the current domain size is expected to be computationally expensive. Therefore, we use a coarse-graining method to accelerate the simulation. Specifically, this coarse-graining method was developed for the modified JKR model in a dense particle medium system with polydisperse particles \citep{kazidenov2023coarse}. Moreover, it has been successfully validated with the same PSD and material parameters used in this work. The material parameters and PSD used in this simulation are provided in Table \ref{tab:mat_prameters} and Table \ref{tab:psd}, respectively. 

\begin{table}[H]
\fontsize{10}{12}\selectfont \caption{Input material parameters of particles used in the simulation.}
\begin{centering}
    \begin{tabular}{ p{5cm} p{2cm}}
\toprule 
Parameter \\
\midrule
 Density ($\mathrm {kg/m^{3}}$)      & 2605 \\
 Young's modulus ($\mathrm {Pa}$)   & $2\cdot10^{10}$   \\
 Surface energy density ($\mathrm {J/m^{2}} $)  & 60  \\
 Friction coefficient  & 0.2   \\
 Restitution coefficient  & 0.8   \\
 Poisson's ratio   & 0.3   \\
 Number of particles  & 33 750    \\
\bottomrule
\end{tabular}
  \label{tab:mat_prameters}
\par\end{centering}
\end{table}

\begin{table}[H]
\fontsize{10}{12}\selectfont \caption{Particle size distribution used in the numerical model.}
\begin{centering}
\begin{tabular}{ p{2.5cm} p{1cm} p{1cm}  p{1cm} p{1.25cm} p{1.25cm} p{1.25cm} p{1cm} p{1cm} }
\toprule
 Diameter (mm)    & 0.3 & 0.36 & 0.4 & 0.44 & 0.5 & 0.55 & 0.6 & 0.71\\

\hline 
Mass fraction     & 0.058  & 0.075 &  0.088  & 0.1216 & 0.2264 & 0.1705 & 0.121 & 0.1395 \\
\bottomrule
\end{tabular}
  \label{tab:psd}
\par\end{centering}
\end{table}



As the particles are generated, the numerical sample is compacted to achieve the desired shape and porosity, which are consistent with the experiment. The porosity of the numerical sample is equal to 40 \%. The next step includes the confinement process, in which the sample is compressed with confining stress of 500 psi. It should be noted that the back and front planes are in a fixed position, and the sample is confined by only side planes. In this stage, we use the modification of the JKR contact model \citep{rakhimzhanova2019numerical} instead of the Hertz model \citep{hertz1882ueber} to initiate the cementation process of the numerical sample and examine the bonding behavior of the particles. In the modified JKR model, the surface energy density determines how strongly the particles are bonded between each other and represent the cementation level of the sample in the experiment. We determine the bonding behavior by the total bonds in the sample. 

\begin{figure}[H]
\begin{centering}
\includegraphics[width=0.6\columnwidth]{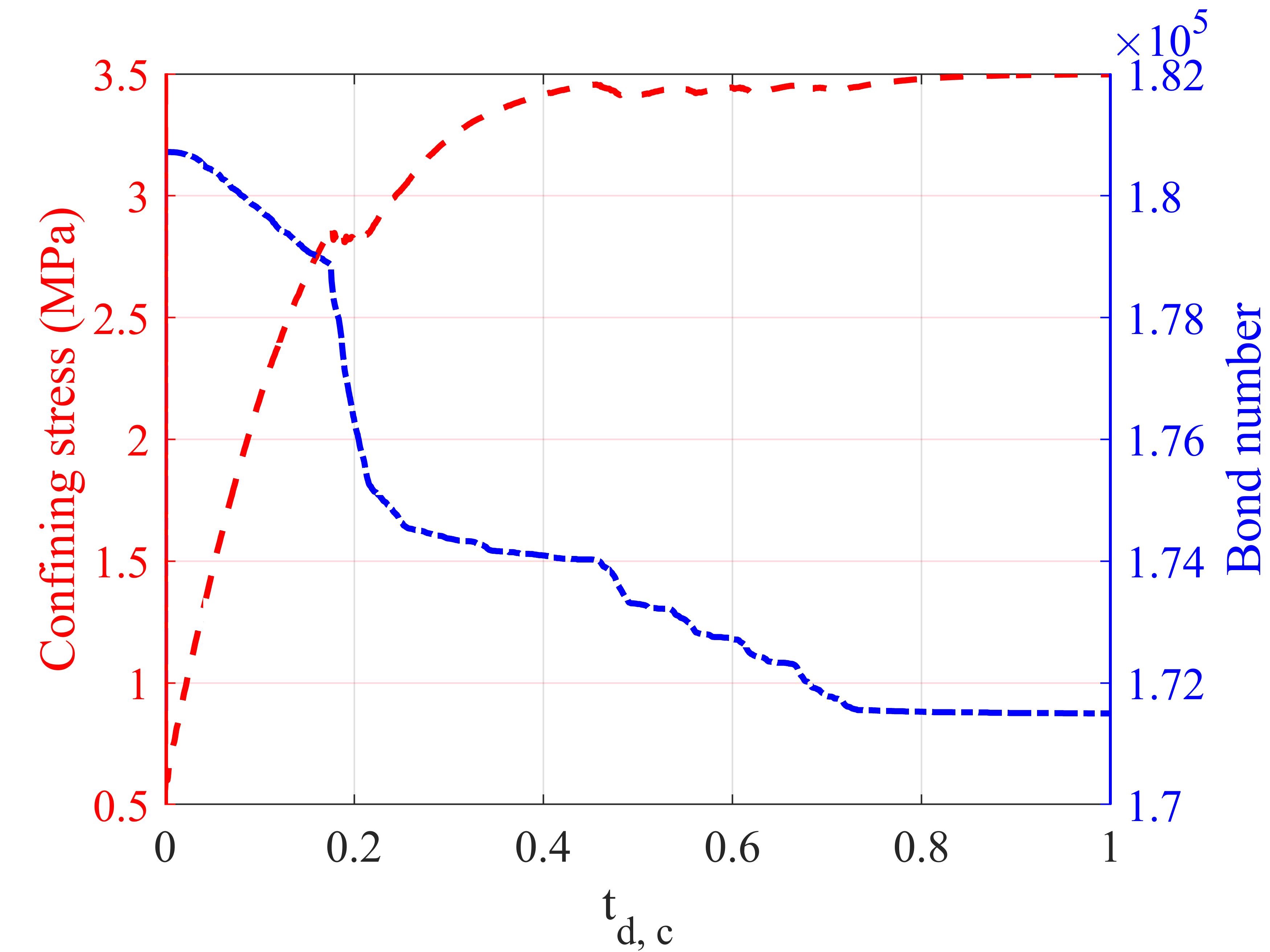}
\par\end{centering}
\caption{Confining stress and total bond number in the sample.\label{fig:Stress_and_bond}}
\end{figure}

Figure \ref{fig:Stress_and_bond} shows the average confining stress and total bond number in the sample during the confinement process. $t_{d, c}$ is the dimensionless time that expresses the duration of the confinement process. We observe that the total number of bonded particles decreases as the stress applied to the sample increases. During the elastic deformation, which lasts from $t_{d, c} = 0$ to about $t_{d, c}  = 0.2 $, the number of bonds reduces moderately from $1.81 \cdot 10^{5} $ to $1.79 \cdot 10^{5} $. At the yield point ($t_{d, c} = 0.2 $), there is an intensive bond breakage, which results in a dramatic decrease in bond number from $1.79 \cdot 10^{5} $ to $1.75 \cdot 10^{5} $ during the short period. Then, we experience the gradual breakage of bonds for almost half of the simulation time, in which the total number of bonds reaches the value of $1.715 \cdot 10^{5} $ at about $t_{d, c} = 0.8 $. As the confining stress becomes stable, the number of bonds remains unchanged.

\begin{figure}[H]
\begin{centering}
\includegraphics[width=1\columnwidth]{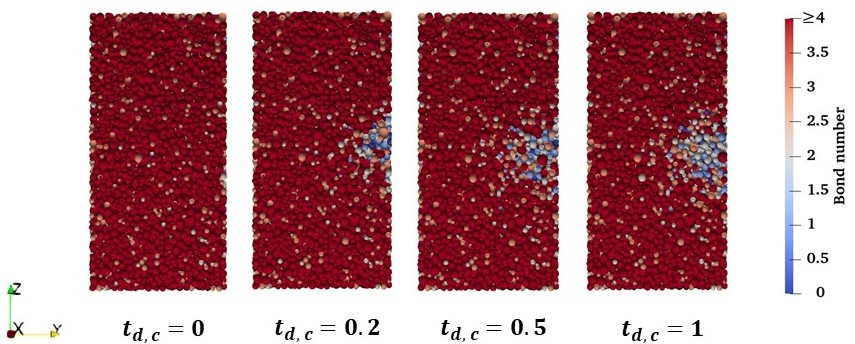}
\par\end{centering}
\caption{Number of bonds per particle during the confinement process.\label{fig:bond_conf}}
\end{figure}

Figure \ref{fig:bond_conf} demonstrates the snapshots of the confining process at different times. The color bar shows the bond number per particle, in which red is a particle with a maximum number of bonds and blue is an unbonded particle. Due to the compression, the bond breakage occurs near the outlet hole, which is then produced. We notice the high magnification of the area of broken particles with increasing stress.

\begin{figure}[H]
\begin{centering}
\includegraphics[width=0.7\columnwidth]{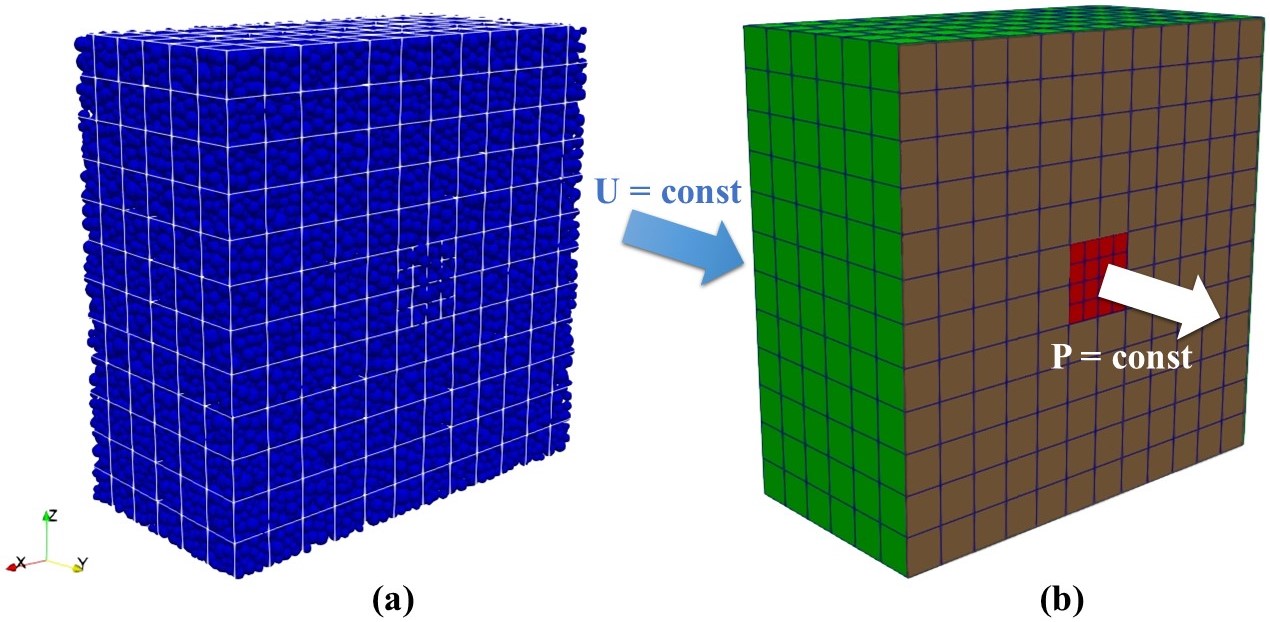}
\par\end{centering}
\caption{(a) Meshing and (b) boundary conditions of the CFD-DEM coupling model.\label{fig:BC}}
\end{figure}

After some time, as the confining stress becomes stable the production of sand stops. Then, we the coupling process of DEM and CFD to model the sand production by polymer flooding. The DEM geometry is fixed at this stage, and CFD geometry is built to correspond to the particle sample with the same dimensions. Therefore, the position and dimensions of the CFD geometry are the same as in the DEM part. Figure \ref{fig:BC} shows the meshing and boundary conditions of the CFD model. In Figure \ref{fig:BC}a, to handle the unresolved case \citep{clarke2018investigation}, the CFD domain is segmented into a grid of 12x6x12 cells along the x, y, and z axes accommodating the multiple particles within the cell. The unresolved case is described below by the following equation:
  \begin{equation}
 \frac{\Delta h}{\bar{d_p}} > 3 \label{eq:13}
 \end{equation}
 where $\Delta h$ is the size of a single fluid cell and  $\bar{d_p}$ is the particle average diameter. 

The size of the simulation time-step significantly impacts the stability and accuracy of the numerical modeling. Therefore, selecting the optimal time-step size is essential to obtain reliable and precise results. Smaller time-steps typically produce more precise results but also result in longer simulation runtimes. On the other hand, using a longer time-step may lead to unpredictable and unrealistic results. The Rayleigh time-step \cite{li2005comparison} is one of the common methods for selecting the most appropriate time-step size for DEM that ensures accurate and stable simulations. The Rayleigh critical time-step is expressed as follows:
  \begin{equation}
\Delta t_{c} = \frac{\pi \bar{R}}{\beta} \sqrt{\frac{\rho_p}{G}} \label{eq:Rayleigh}
\end{equation}

where $\bar{R}$ is the average radius of the particle, $\rho_p$ is the density of the particles, $G = \frac{E}{2(1-\nu)}$ is the shear modulus of the particle, where $\nu$ is the Poisson's ratio, and $\beta = 0.8766 + 0.163 \nu$.

In the CFD domain, the stability of the system is determined by the flow propagation time across the single fluid cell. Physically, the flow time through the cell should be at most one time step for correct operation. We use the Courant-Friedrichs-Lewy (CFL) condition \citep{courant1928partiellen} to calculate the critical time-step for CFD as follows:
\begin{equation}
C=\frac{U \Delta t_{CFD}}{\Delta h}<C_{max} \label{eq:CFL}
\end{equation}
where $C$ is the Courant number, $U$ is the fluid flow velocity and $\Delta t_{CFD}$ is the time-step of the CFD simulation. The value  of $C_{max}$ depends on the solver time-integration scheme. Typically, it should be much less than 1 to maintain an ideal stable system.


The DEM and CFD time-steps in the coupling simulations may operate in consecutive or concurrent regimes \citep{kloss2012models}. Data interchange between the DEM and CFD takes place at the same core sequentially in the consecutive regime. In this situation, it is best to use resources efficiently by allowing all cores to be active at all times. In the concurrent regime, the DEM and CFD coupling computations occur parallel at the same time-step using separate cores. The time-step difference between the two unique models affects how stable CFD-DEM coupling simulations operate. By choosing the appropriate CFD and DEM time-steps, one may manage the simulation duration and enhance the precision of the results. The DEM time-step is often much smaller than the CFD time-step. For every CFD time-step, there might be several numbers of DEM time-steps. In this simulation, we choose the case in which the DEM time-step of $\Delta t_{DEM} = 5 \cdot 10^{-8} s $ is 100 times smaller than the CFD time-step, which is equal to $\Delta t_{CFD} = 5 \cdot 10^{-6} s $. These time-step options are adapted from the work of \cite{kazidenov2023time}, where they successfully found and proved the optimal time-step selection for CFD and DEM by investigating the sand production in the Kazakhstan oilfield reservoir. According to equation \ref{eq:Rayleigh}, the DEM time-step size, which is 4.26 \% of the Rayleigh critical time-step, is within the acceptance range. The CFD time-step is also satisfied the CFL condition (equation \ref{eq:CFL}) with Courant number of $5.38 \cdot 10^{-7}$, which is substantially smaller than 1.

In the CFD part, Xanthan Gum (XG) polymer solution, selected as an injection fluid, is characterized as a non-Newtonian fluid. In the simulation, we use the power-law parameters from Table \ref{tab:PL_param}, obtained by investigating its rheology. Figure \ref{fig:BC}b shows the boundary conditions of the CFD. The side planes to which the confining stress is applied are assigned in periodic boundary conditions depicted in green. The front plane is in "no slip", represented by the brown color, while the outlet hole corresponds to an atmospheric pressure of $P = 0$ (red color). We inject the fluid into the domain from the backplane, opposite the front plane. The injection velocity of the fluid is $U = 1.55 \cdot 10^{-4}$ m/s, which corresponds to the laboratory experiment. The initial conditions are $U(0) = 0$ and  $P(0) = 0$. The rectangular outlet hole in the CFD is resolved into smaller cells to match the dimensions of the circular hole in the DEM.

\section{Results and discussion}
\label{sec:results}
\subsection{Experimental results}

\subsubsection{Brine injection}
\begin{figure}[H]
\begin{centering}
\includegraphics[width=0.6\columnwidth]{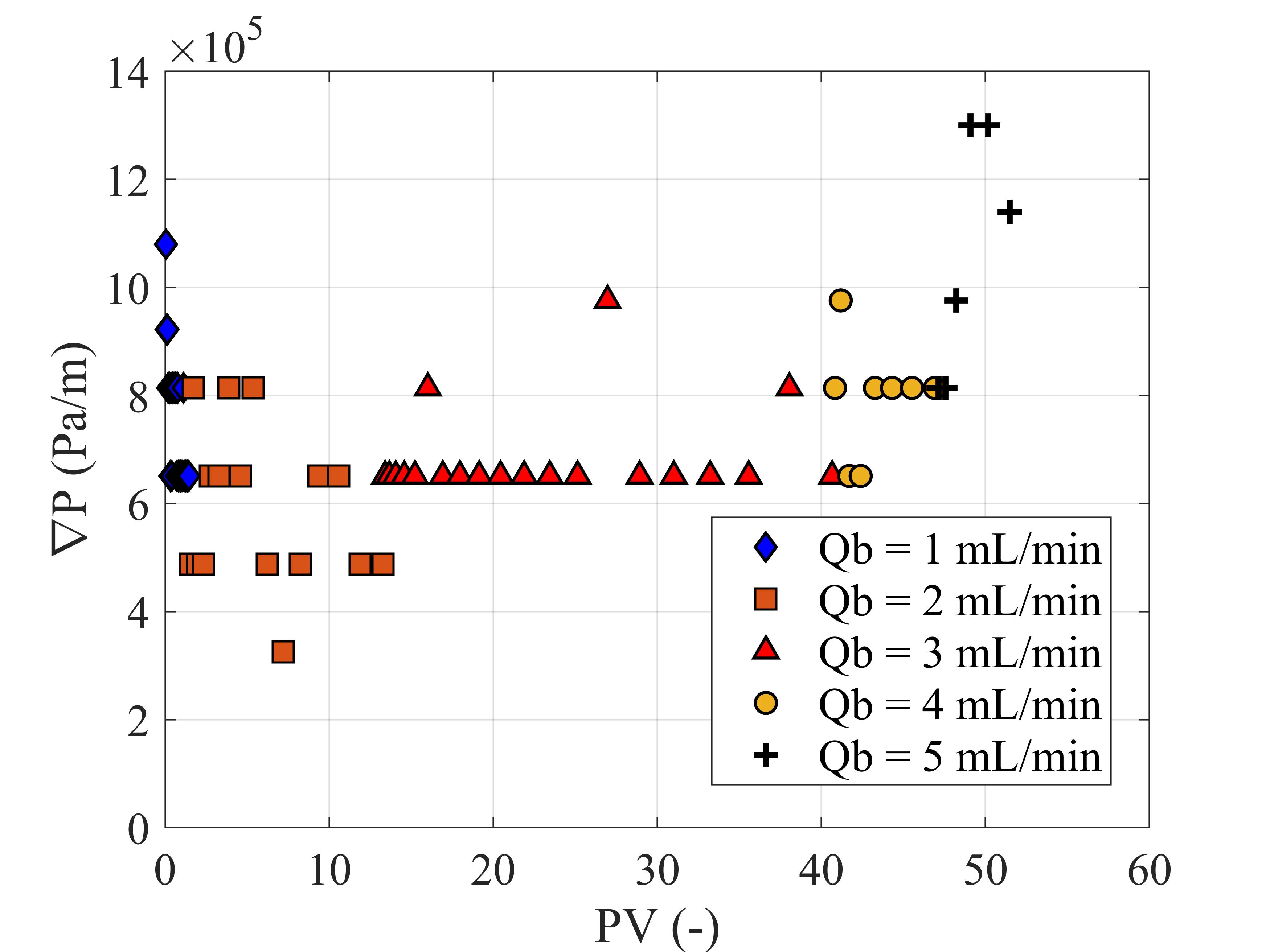}
\par\end{centering}
\caption{Pressure gradient as a function of PV at various flow rates.\label{fig:brine_inj}}
\end{figure}

Figure \ref{fig:brine_inj} shows the pressure gradient versus injected pore volume of brine at different flow rates. We observed a general trend of increasing pressure gradient with flow rate. Notable fluctuations in the pressure gradient were seen at flow rates of 1 and 2 mL/min and this is because the pressure sensors were not accurate enough at low flow rates. Nevertheless, the pressure transmitters showed steady results of $6.51 \cdot 10^{5}$ Pa/m from 13 to 40 PV of injection at a flow rate of 3 mL/min. At a flow rate of 4 mL/min, the pressure gradient increases up to $8.14 \cdot 10^{5}$ Pa/m, and a value of $13.00 \cdot 10^{5}$ Pa/m is observed at the flow rate of 5 mL/min. The impact of brine injection on sand production was also studied from the effluent at the outlet. However, sand production was not observed at all flow rates.

\subsubsection{Impact of polymer flooding on sand production}
After the brine injection test, and determining the permeability of the core (see Table \ref{tab:pm_parameters}), the XG polymer solution is injected horizontally. The flow rates below 3 mL/min were not studied because of the instability of pressure sensors at lower flow rates. As mentioned before, the XG solution is a non-Newtonian shear-thinning fluid whose rheogram is presented in Figure \ref{fig:XG}. 

\begin{figure}[htbp]
\begin{centering}
\includegraphics[width=0.6\columnwidth]{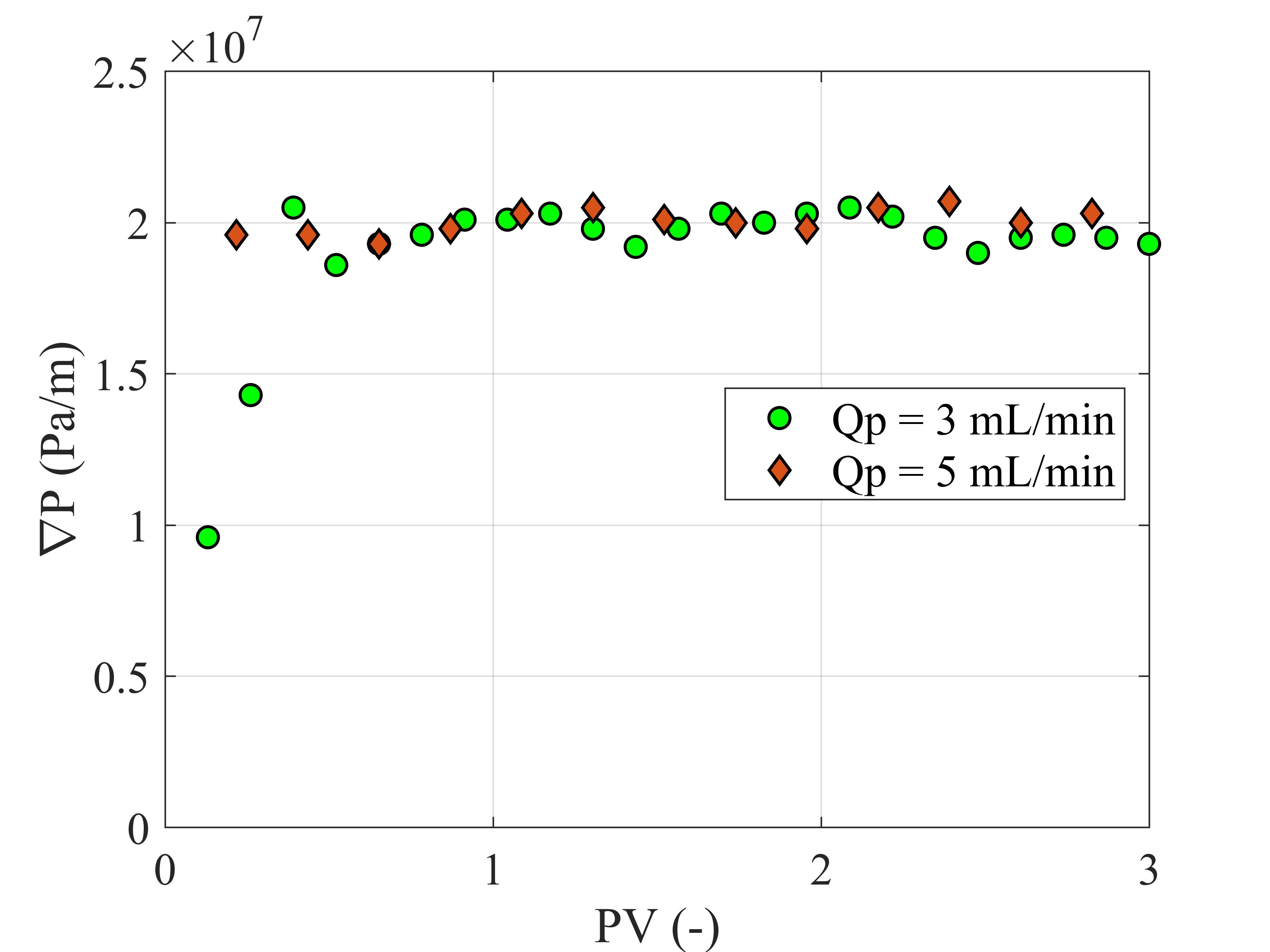}
\par\end{centering}
\caption{Pressure gradient as a function of injected PV of polymer solution at the flow rate of 3 and 5 mL/min.
\label{fig:dif_flow_rate_inj}}
\end{figure}

Figure \ref{fig:dif_flow_rate_inj} presents the evolution of the pressure gradient per pore volume of polymer injected (in PV) at 3 and 5 mL/min flow rates. The breakthrough of the polymer solution occurred around 0.82 PV for both flow rates. We noticed that the pressure gradient is the same, around $2.00 \cdot 10^{7}$ Pa/m, despite the change in flow rate from 3 to 5 mL/min. This phenomenon can be explained by the non-Newtonian behavior of the polymer solution because the viscosity of the XG solution decreases by increasing the flow velocity, i.e., the shear rate (see Figure \ref{fig:XG}). 

\begin{figure}[H]
\begin{centering}
\includegraphics[width=0.6\columnwidth]{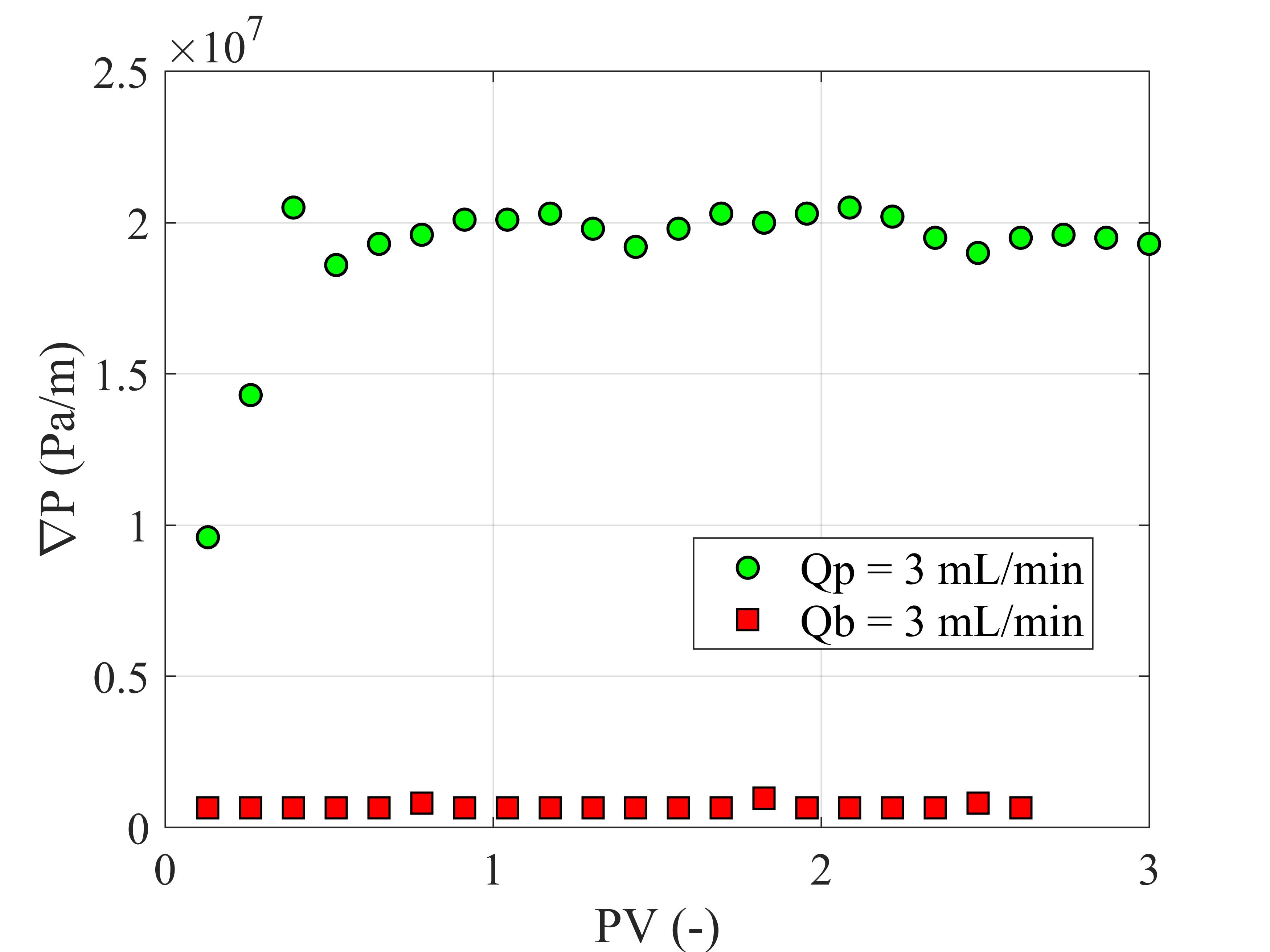}
\par\end{centering}
\caption{Pressure gradient vs PV of polymer and brine injection at 3 mL/min.\label{fig:brine_vs_XG}}
\end{figure}

To observe the impact of polymer on the pressure gradient, we plotted the pressure gradient against PV data of the brine and XG polymer solution in Figure \ref{fig:brine_vs_XG}. The polymer solution increased the pressure gradient by a factor of 30 compared to the brine injection. Assuming that the equivalent shear rate is at the maximum value of 100 1/s, the viscosity is roughly 0.08 Pa.s, which is 80 times more viscous than water. However, note that the XG solution viscosity field distribution across the core will differ since the shear rate distribution can vary across the core sample due to the pore radii.

\begin{figure}[H]
\begin{centering}
\includegraphics[width=0.6\columnwidth]{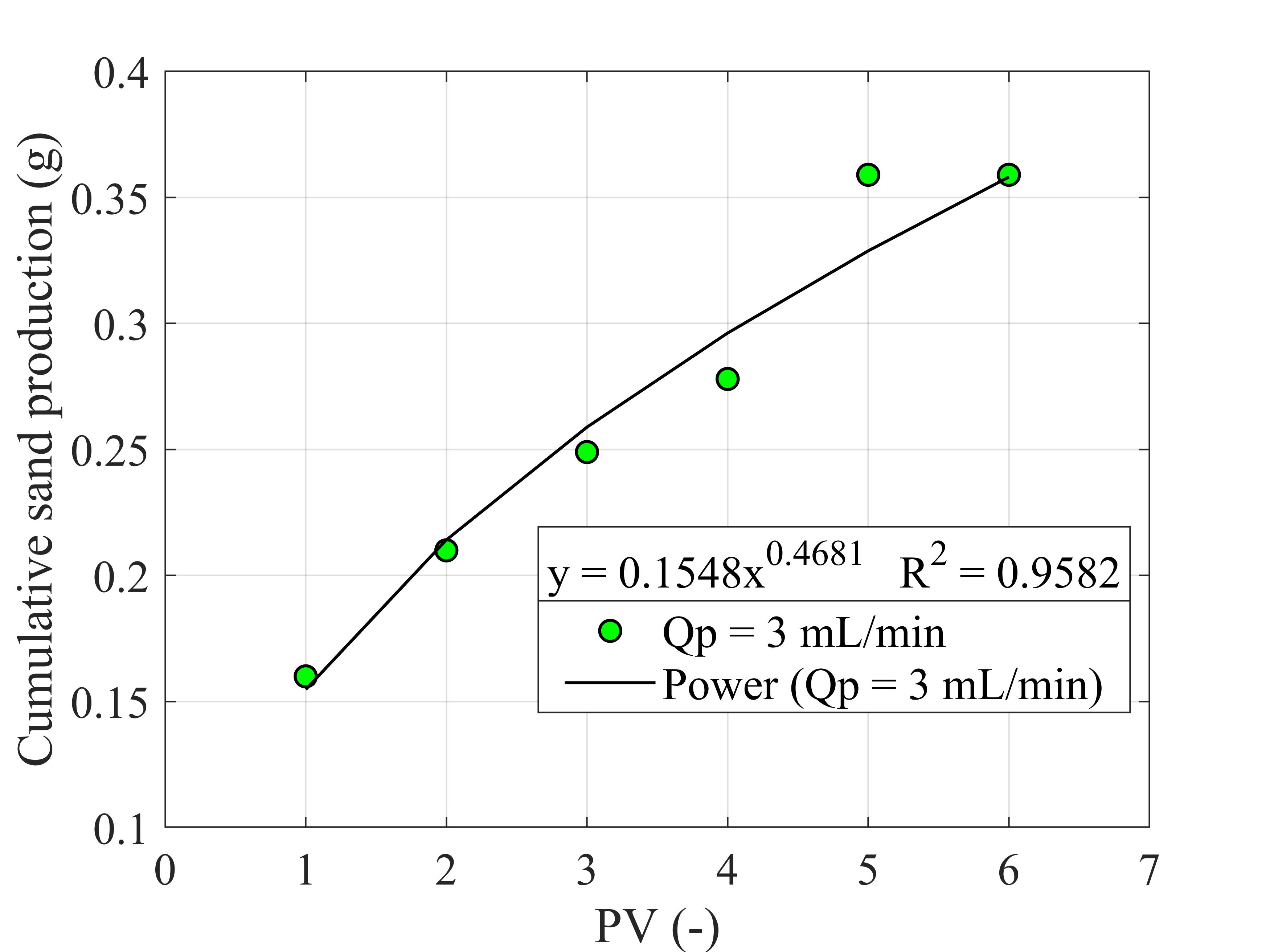}
\par\end{centering}
\caption{Cumulative sand production as a function of PV at Q=3 mL/min.\label{fig:sand_prod_exp}}
\end{figure}

As was mentioned in the previous section, there was no sand production during the brine injection, even at the flow rate of 5 mL/min.  However, we noticed sand production for the polymer injection at a flow rate of 3 mL/min, plotted in Figure \ref{fig:sand_prod_exp}. Due to the higher viscosity (i.e., viscous forces), the XG polymer solution mobilized the sand particles. Moreover, a solid shear-thinning behavior of XG solution may enhance the mobilization of poorly consolidated sand particles. 
As a result,  a total of 0.359 g of sand was produced after 5 PV of injection of the polymer solution. 44.5\% of total sand is produced after 1 PV of polymer injection, and sand production decreases with injected PV.Based on the trend equation on the results shown in the figure, the cumulative sand production equals $0.15PV^{0.47}$.

\subsection{Numerical results}

Initially, the numerical model is verified with the experimental data by comparing the dimensionless cumulative sand production rate along the dimensionless time, which is given as follows:
\begin{equation}
t_{d} = \left(\displaystyle \frac{t}{t_{end}}\right)\label{eq:39}
\end{equation}
where $t$ is the current time, $t_{end}$ is the transient phase starting time when the cessation of sand production occurs. We define the dimensionless cumulative sand production rate using the following expression:
\begin{equation}
M^{t}_{d} = \left(\displaystyle \frac{ \int_{t=0}^{t}M_{t}dt}{\int_{t=0}^{t_{end}}M_{t}dt}\right)\label{eq:40}
\end{equation}
where $M_{t}$ is the cumulative mass of produced sand at time $t$. 

\begin{figure}[H]
\begin{centering}
\includegraphics[width=0.6\columnwidth]{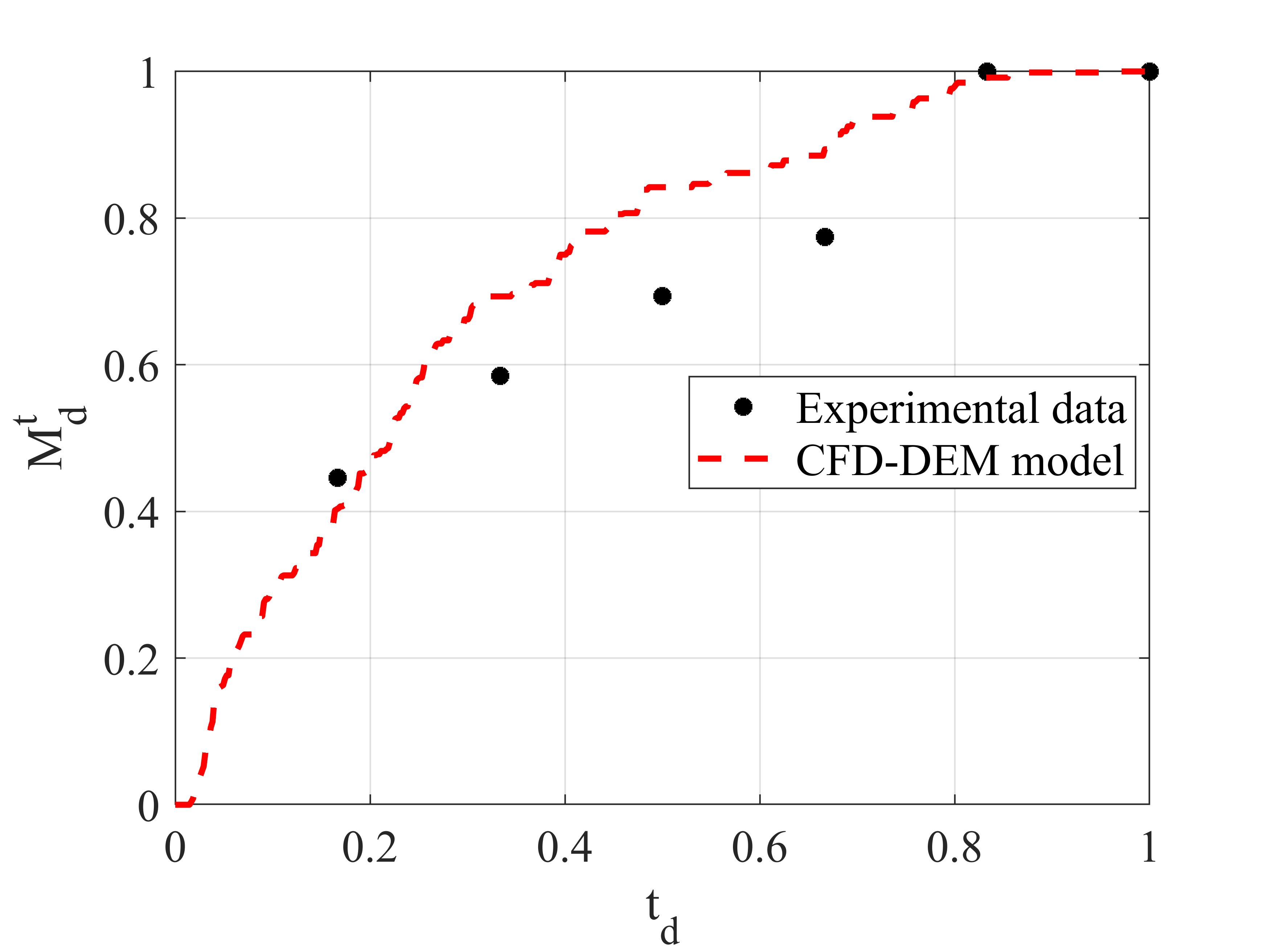}
\par\end{centering}
\caption{Comparison of dimensionless cumulative sand production rate of the experiment and numerical simulation.\label{fig:fig1}}
\end{figure}

Figure \ref{fig:fig1} compares the dimensionless cumulative mass of produced sand of the numerical model and experimental data at time $t_{d}$. The red curve, which results from numerical simulation, shows only the sand production due to fluid injection. The mass of the sand produced during the confinement process is not considered to examine only the polymer effect on sand production. Although the experimental results contain only six data, it clearly reflects the general pattern of sand production. From the last two points, it can be concluded that sand production during this period remains unchanged and acquires a transitional character. We observe that the numerical model results are in relatively good agreement with the experimental data and demonstrate a similar pattern of the curve in dimensionless cumulative sand production rate.

To better examine the accuracy between the numerical model and experimental data, the root mean squared relative error (RMSRE) is calculated using the following equation:
\begin{equation}
    RMSRE =\sqrt{ \frac{1}{n} \sum_{i=1}^{n} \Bigg| \frac{M^i_{d, exp} - M^i_{d,model}}{M^i_{d, exp}}}\Bigg|^{2} 
    \label{eq:eq14}
\end{equation}
where $M_{d,{exp}}^i$ is the $i$-th dimensionless cumulative produced sand mass in the experiment and $M_{d,{model}}^i$ is the $i$-th dimensionless cumulative produced sand mass in the model. In accordance with equation (\ref{eq:eq14}), the RMSRE value for the numerical model is 0.137.  

\begin{figure}[htbp]
\begin{centering}
\includegraphics[width=1\columnwidth]{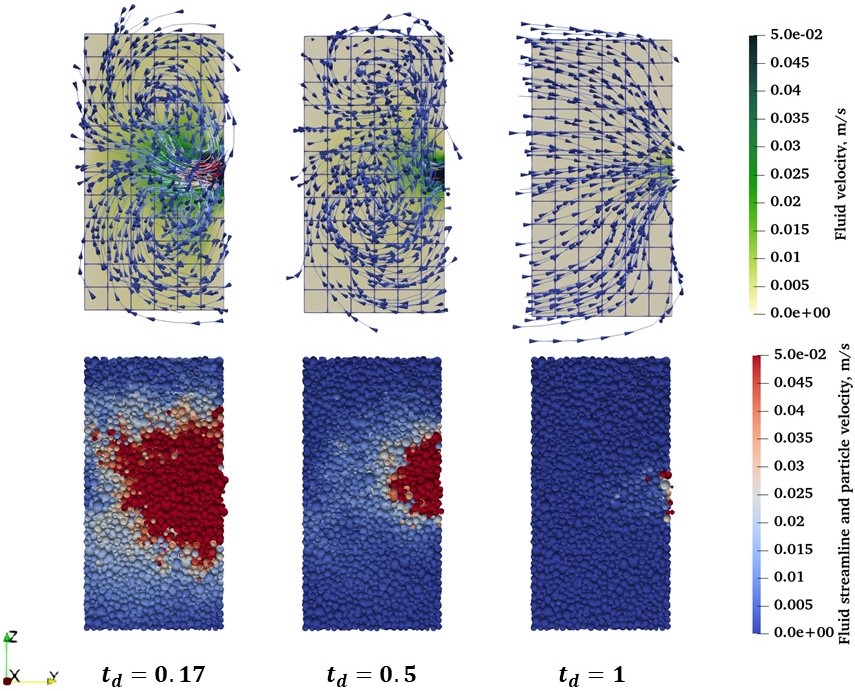}
\par\end{centering}
\caption{Snapshots of fluid velocity, streamline and particle velocity at different time periods.\label{fig:Streamlines}}
\end{figure}

Figure \ref{fig:Streamlines} represents a series of snapshots of the sample's fluid velocity, fluid streamline, and particle velocities. The capture time of the snapshots corresponds to the 1$^{st}$, 3$^{rd}$ and 6$^{th}$ points of the experimental data, which are $t_d = 0.17$, $t_d = 0.5$, and $t_d = 1$ in dimensionless time, respectively. These snapshots provide clear insights into fluid flow behavior and particle movement at different sand production phases. For example, at the beginning of injection, the fluid has a higher velocity at the center near the outlet hole with a non-uniform streamlines distribution, leading to a greater velocity of sand particles. Therefore, we observe a higher rate of sand production due to the intensive movement of sand toward the outlet. However, as time progresses, fluid and particle velocity decreases, resulting in reduced sand production. At the final time, the velocity of the fluid substantially decreases with uniformly distributed streamlines. The sand production enters the transient regime, and no further production occurs. 

\begin{figure}[htbp]
\begin{centering}
\includegraphics[width=1\columnwidth]{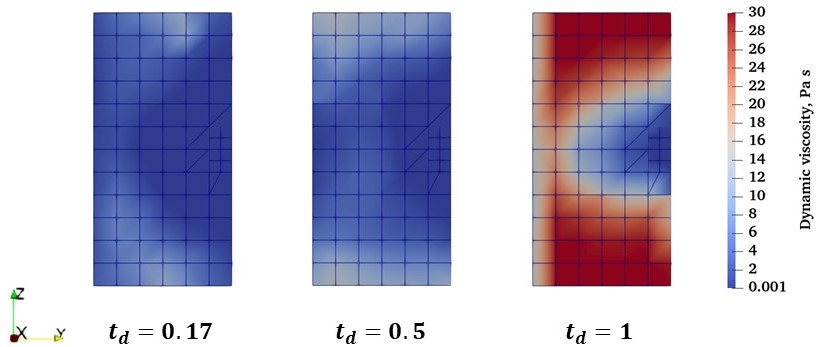}
\par\end{centering}
\caption{Viscosity of the fluid in the sample at different time periods.\label{fig:visc}}
\end{figure}

The fluid flow velocity in the sample mainly influences the fluid's viscosity. Figure \ref{fig:visc} demonstrates the change in apparent fluid viscosity in the sample at different periods. During the intensive sand production  ($t_d = 0.17$) and at the midpoint ( $t_d = 0.5$),  the fluid exhibits lower viscosity at the whole sample. As flow is uniformly distributed throughout the sample, the fluid becomes more viscous in more tightly bonded regions where the flow velocity is less. On the other hand, in areas with unbonded particles, especially near the outlet hole, the fluid viscosity decreases. 

\begin{figure}[H]
\begin{centering}
\includegraphics[width=0.6\columnwidth]{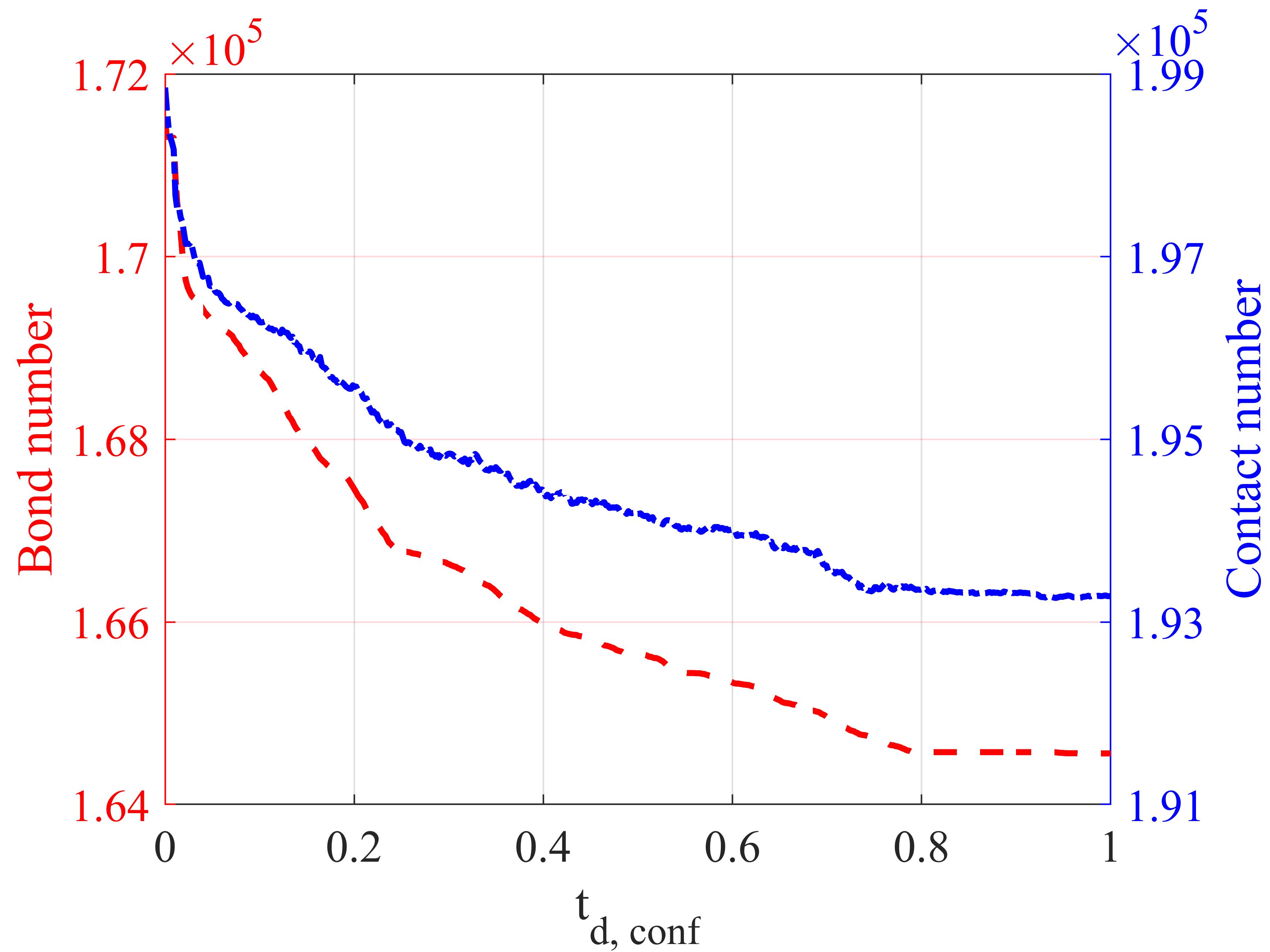}
\par\end{centering}
\caption{Total number of bonds and contacts between the particles in the sample during the polymer injection.\label{fig:Bond_and_contact}}
\end{figure}

Figure \ref{fig:Bond_and_contact} demonstrates the total number of bonds and contacts between particles in the sample. We see that the injected polymer also affects bond breakage. The fluid not only flushes out the already broken particles due to the confinement process but also causes the breakage of bonds with their consecutive production. During the confinement process, as shown in Figure \ref{fig:Stress_and_bond}, the bond breakage occurs with some fluctuations due to the applied confining stress and characteristics of the sample material. On the other hand, the injection of a polymer leads to a stable bond breakage with a gradual decrease in the total number of bonds. As the sand production becomes transient, the total number of bonds reduces by $0.8 \cdot 10^{5} $ bonds from the initial injection state, reaching the $1.645 \cdot 10^{5} $ bonds in the sample. During the confinement process, the reduction of the bonds is equal to $0.9 \cdot 10^{5} $, which is slightly higher than in polymer injection. These findings indicate that the polymer injection demonstrates relatively similar results to the confinement process regarding bond breakage. 

The blue curve in Figure \ref{fig:Bond_and_contact} displays the total number of contacts in the sample during the polymer injection. Initially, the contacts decrease sharply from $1.99 \cdot 10^{5}$ to $1.97 \cdot 10^{5}$ and demonstrate a similar curve pattern to the curve of the bond number. This implies that the polymer with a higher initial velocity effectively segregates all unbonded particles. However, as the fluid becomes steady, the contacts reduce less compared to the bonds, indicating that some particles remain in contact even though they do not adhesively interact with each other. After the gradual reduction, both the bond and contact number reach the constant value at approximately $t_{d} = 0.8$ dimensionless time, which shows the cessation of sand production. 


\begin{figure}[H]
\begin{centering}
\includegraphics[width=1\columnwidth]{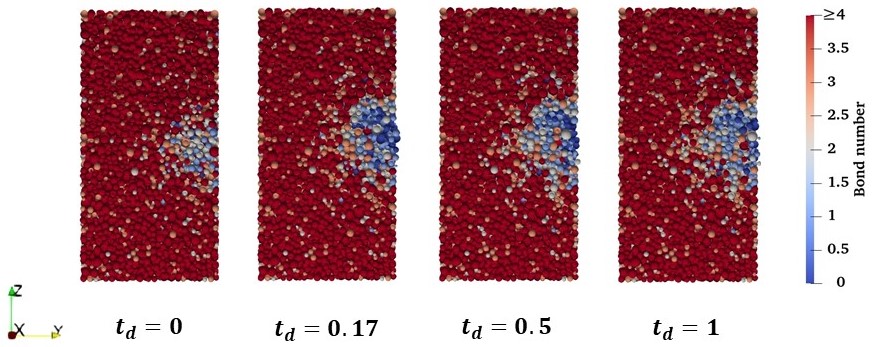}
\par\end{centering}
\caption{Bond number per particle at different time periods.\label{fig:Snap_bonds}}
\end{figure}

Figure \ref{fig:Snap_bonds} shows the snapshots of the bond number per particle during the polymer injection. The snapshots are captured at $t_d = 0$, $t_d = 0.17$, $t_d = 0.5$ and $t_d = 1$. While there is a slight increase in the breakage area between  $t_d = 0$ and $t_d = 0.17$, we do not see significant differences from $t_d = 0.17$ to $t_d = 1$ since the bonds reduce gradually at this period. Moreover, it should be considered that some of the broken particles leave the sample through the hole as time progresses. Therefore, this effect contributes to the similar area of broken particles near the hole. 

\begin{figure}[H]
\begin{centering}
\includegraphics[width=0.6\columnwidth]{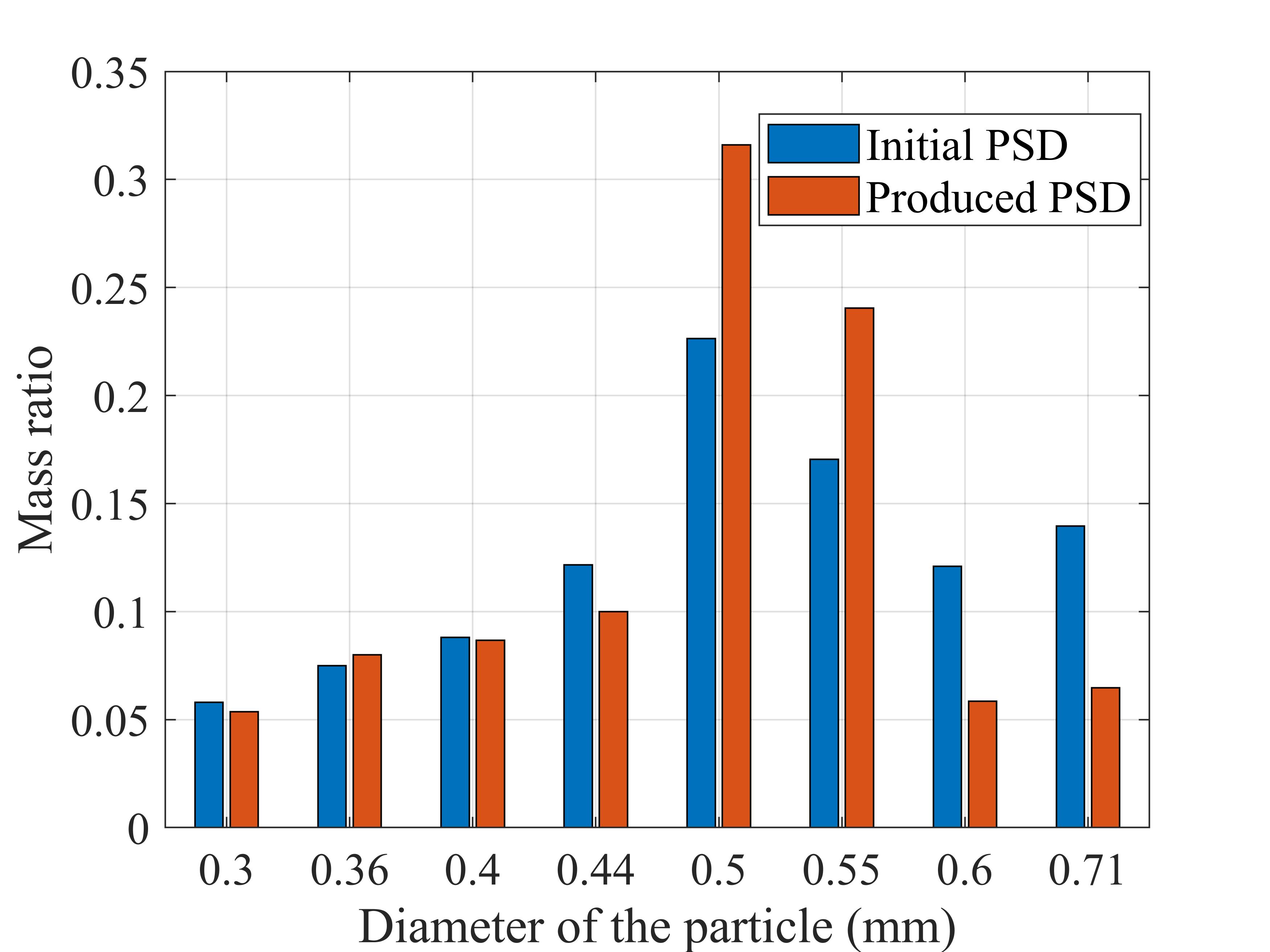}
\par\end{centering}
\caption{PSD of the initial and produced particles.\label{fig:PSD_produced}}
\end{figure}

Since the sample contains particles of different sizes, we compare the size distribution of the produced particles with the initial PSD of the sample. Figure \ref{fig:PSD_produced} compares the mass ratio of each particle size of the initial and produced particles. The final PSD of the produced particles is measured when the sand production becomes in the transient phase. We notice that the mass of the small particles is less differentiated than that of the large ones. For example, the first four sizes of particles (0.3 mm, 0.36 mm, 0.4 mm, and 0.44 mm), which are smaller in size, have approximately similar mass ratios of initial and produced particles. While the mass ratio of 0.5 and 0.55 mm produced particles is more than that in initial particles, the mass ratio of 0.6 and  0.71 mm produced particles is less than the mass ratio of the identical size particles in the sample before fluid injection. These findings suggest that the significant source of sand production can be attributed to particles of 0.5 mm and 0.55 mm.

\section{Conclusions}
\label{sec:conclusion}

The primary aim of this study is to experimentally and numerically investigate the effect of polymer injection on sand production in poorly consolidated sandstone. The artificially made sandstone is prepared based on the PSD of the Kazakhstan oil field. The production of sand from this sandstone is studied by injecting the polymer solution whose viscosity has been characterized using rotational rheometry. In the numerical model, the modified cohesive contact model is used to model the cementation behavior of the sample. The power-law model describes the non-Newtonian fluid flow. The fluid-particle interaction is characterized by solving the drag force based on the power-law model in the CFD and DEM coupling.  

Experimentally no sand was produced by injecting brine solution, even at high flow rates at a constant radial confining pressure of 500 psi. In contrast, we observed the impact of the polymer solution injection on the production of sand. Moreover, 44.5\% of produced sand was obtained after 1 PV of polymer solution injection, i.e., breakthrough. The power-law model well represents the Xanthan gum solution, and the fitted parameters were used as input for numerical investigation.

The numerical model is verified with the results of the laboratory experiment. The findings suggest that the numerical model is consistent with the experimental data demonstrating a similar curve pattern in the dimensionless cumulative sand production rate by fluid injection. To examine the sanding behavior by polymer flooding, we compare different results such as fluid and particle velocities, fluid viscosity, bonding behavior of sand particles, and PSD of produced particles at different time intervals. We observe that the polymer injection has a significant effect on bond breakage. While the bonded particles break in an unstable manner during the confinement process, the unbonding of the particles occurs gradually during the fluid injection. Therefore, sand production happens not only from already broken particles due to the confinement but also due to polymer injection. At the beginning of the injection, we noticed an intensive sand production rate caused by the higher velocity of the fluid. The sand production rate enters the transient phase as the fluid flow becomes stationary. The fluid has a lower viscosity in the higher permeable zones caused by dramatic unbonding. On the other hand, the fluid viscosity is higher in regions with an abundance of bonded particles. The mass ratio difference between produced particles and the initial sample is much more significant for larger particles. The fine particles have approximately the same mass ratios. Almost 60 \% of the produced sand consists of medium-sized particles with a diameter of 0.5 mm and 0.55 mm. 

This knowledge can guide the study of sand production by polymers, i.e., non-Newtonian fluids in porous media, especially for poorly consolidated porous media. We expect our study to be a starting point for further research on sand production by non-Newtonian fluids in poorly consolidated porous media.


\bibliographystyle{plainnat}
\bibliography{references}  

\begin{thebibliography}{54}
\providecommand{\natexlab}[1]{#1}
\providecommand{\url}[1]{\texttt{#1}}
\expandafter\ifx\csname urlstyle\endcsname\relax
  \providecommand{\doi}[1]{doi: #1}\else
  \providecommand{\doi}{doi: \begingroup \urlstyle{rm}\Url}\fi

\bibitem[Atapattu et~al.(1995)Atapattu, Chhabra, and
  Uhlherr]{atapattu1995creeping}
DD~Atapattu, RP~Chhabra, and PHT Uhlherr.
\newblock Creeping sphere motion in herschel-bulkley fluids: flow field and
  drag.
\newblock \emph{Journal of non-newtonian fluid mechanics}, 59\penalty0
  (2-3):\penalty0 245--265, 1995.

\bibitem[Bautista and Dahi~Taleghani(2017)]{bautista2017state}
JF~Bautista and A~Dahi~Taleghani.
\newblock The state of the art and challenges in geomechanical modeling of
  injector wells: a review paper.
\newblock \emph{Journal of Energy Resources Technology}, 139\penalty0 (1),
  2017.

\bibitem[Clarke et~al.(2018)Clarke, Sederman, Gladden, and
  Holland]{clarke2018investigation}
Daniel~A Clarke, Andrew~J Sederman, Lynn~F Gladden, and Daniel~J Holland.
\newblock Investigation of void fraction schemes for use with cfd-dem
  simulations of fluidized beds.
\newblock \emph{Industrial \& Engineering Chemistry Research}, 57\penalty0
  (8):\penalty0 3002--3013, 2018.

\bibitem[Courant et~al.(1928)Courant, Friedrichs, and
  Lewy]{courant1928partiellen}
Richard Courant, Kurt Friedrichs, and Hans Lewy.
\newblock {\"U}ber die partiellen differenzengleichungen der mathematischen
  physik.
\newblock \emph{Mathematische annalen}, 100\penalty0 (1):\penalty0 32--74,
  1928.

\bibitem[Cundall and Strack(1979)]{cundall1979discrete}
Peter~A Cundall and Otto~DL Strack.
\newblock A discrete numerical model for granular assemblies.
\newblock \emph{geotechnique}, 29\penalty0 (1):\penalty0 47--65, 1979.

\bibitem[Dazhi and Tanner(1985)]{dazhi1985drag}
Gu~Dazhi and RI~Tanner.
\newblock The drag on a sphere in a power-law fluid.
\newblock \emph{Journal of Non-Newtonian Fluid Mechanics}, 17\penalty0
  (1):\penalty0 1--12, 1985.

\bibitem[Fang et~al.(2022)Fang, Wang, Zhong, Qiu, and Wang]{fang2022cfd}
Xing Fang, Guorong Wang, Lin Zhong, Shunzuo Qiu, and Dangfei Wang.
\newblock A cfd--dem analysis of the de-cementation behavior of weakly cemented
  gas hydrate--bearing sediments in a hydrocyclone separator.
\newblock \emph{Particulate Science and Technology}, 40\penalty0 (7):\penalty0
  812--823, 2022.

\bibitem[Fang and Tan(2006)]{fang2006dynamic}
Yang Fang and Xiaobo Tan.
\newblock A dynamic jkr model with application to vibrational release in
  micromanipulation.
\newblock In \emph{2006 IEEE/RSJ International Conference on Intelligent Robots
  and Systems}, pages 1341--1346. IEEE, 2006.

\bibitem[Fattahpour et~al.(2012)Fattahpour, Moosavi, and
  Mehranpour]{fattahpour2012experimental}
Vahidoddin Fattahpour, Mahdi Moosavi, and Mahdi Mehranpour.
\newblock An experimental investigation on the effect of rock strength and
  perforation size on sand production.
\newblock \emph{Journal of Petroleum Science and Engineering}, 86:\penalty0
  172--189, 2012.

\bibitem[Frungieri et~al.(2022)Frungieri, Boccardo, Buffo, Karimi-Varzaneh, and
  Vanni]{frungieri2022cfd}
Graziano Frungieri, Gianluca Boccardo, Antonio Buffo, Hossein~Ali
  Karimi-Varzaneh, and Marco Vanni.
\newblock Cfd-dem characterization and population balance modelling of a
  dispersive mixing process.
\newblock \emph{Chemical Engineering Science}, 260:\penalty0 117859, 2022.

\bibitem[Garner et~al.(2018)Garner, Strong, and Zavaliangos]{garner2018study}
Sean Garner, John Strong, and Antonios Zavaliangos.
\newblock Study of the die compaction of powders to high relative densities
  using the discrete element method.
\newblock \emph{Powder Technology}, 330:\penalty0 357--370, 2018.

\bibitem[Giese and Powers(2002)]{giese2002using}
Steven~W Giese and Susan~E Powers.
\newblock Using polymer solutions to enhance recovery of mobile coal tar and
  creosote dnapls.
\newblock \emph{Journal of contaminant hydrology}, 58\penalty0 (1-2):\penalty0
  147--167, 2002.

\bibitem[Guo et~al.(2019)Guo, Li, Kong, Ma, Li, and Wang]{guo2019lessons}
Hu~Guo, Yiqiang Li, Debin Kong, Ruicheng Ma, Binhui Li, and Fuyong Wang.
\newblock Lessons learned from alkali/surfactant/polymer-flooding field tests
  in china.
\newblock \emph{SPE Reservoir Evaluation \& Engineering}, 22\penalty0
  (01):\penalty0 78--99, 2019.

\bibitem[Hall~Jr and Harrisberger(1970)]{hall1970stability}
CD~Hall~Jr and WH~Harrisberger.
\newblock Stability of sand arches: a key to sand control.
\newblock \emph{Journal of Petroleum Technology}, 22\penalty0 (07):\penalty0
  821--829, 1970.

\bibitem[Hertz(1882)]{hertz1882ueber}
Heinrich Hertz.
\newblock Ueber die ber{\"u}hrung fester elastischer k{\"o}rper.
\newblock 1882.

\bibitem[Iskaziyev et~al.(2013)Iskaziyev, Karabalin, and
  Azhgaliyev]{iskaziyev2013complex}
K~Iskaziyev, U~Karabalin, and D~Azhgaliyev.
\newblock Complex study of sedimentary basins is the basis for effective
  forecast of the oil and gas content in new territories.
\newblock \emph{Pet. Anal. J}, 2013.

\bibitem[Ismail et~al.(2021)Ismail, Kuang, and Yu]{ismail2021cfd}
Noor~Ilyana Ismail, Shibo Kuang, and Aibing Yu.
\newblock Cfd-dem study of particle-fluid flow and retention performance of
  sand screen.
\newblock \emph{Powder Technology}, 378:\penalty0 410--420, 2021.

\bibitem[Johnson et~al.(1971)Johnson, Kendall, and Roberts]{johnson1971surface}
Kenneth~Langstreth Johnson, Kevin Kendall, and aAD Roberts.
\newblock Surface energy and the contact of elastic solids.
\newblock \emph{Proceedings of the royal society of London. A. mathematical and
  physical sciences}, 324\penalty0 (1558):\penalty0 301--313, 1971.

\bibitem[Kazidenov et~al.(2022)Kazidenov, Khamitov, and
  Amanbek]{kazidenov2022coarseconf}
Daniyar Kazidenov, Furkhat Khamitov, and Yerlan Amanbek.
\newblock Coarse-graining methods for the modified jkr contact model on a
  triaxial compression test.
\newblock In \emph{56th US Rock Mechanics/Geomechanics Symposium}. OnePetro,
  2022.

\bibitem[Kazidenov et~al.(2023{\natexlab{a}})Kazidenov, Khamitov, and
  Amanbek]{kazidenov2023coarse}
Daniyar Kazidenov, Furkhat Khamitov, and Yerlan Amanbek.
\newblock Coarse-graining of cfd-dem for simulation of sand production in the
  modified cohesive contact model.
\newblock \emph{Gas Science and Engineering}, page 204976, 2023{\natexlab{a}}.

\bibitem[Kazidenov et~al.(2023{\natexlab{b}})Kazidenov, Omirbekov, and
  Amanbek]{kazidenov2023time}
Daniyar Kazidenov, Sagyn Omirbekov, and Yerlan Amanbek.
\newblock Optimal time-step for coupled cfd-dem model in sand production.
\newblock In \emph{Computational Science and Its Applications--ICCSA 2023
  Workshops: Athens, Greece, July 3--6, 2023, Preprint}. Springer,
  2023{\natexlab{b}}.

\bibitem[Khamitov et~al.(2021)Khamitov, Minh, and Zhao]{khamitov2021coupled}
Furkhat Khamitov, Nguyen~Hop Minh, and Yong Zhao.
\newblock Coupled cfd--dem numerical modelling of perforation damage and sand
  production in weak sandstone formation.
\newblock \emph{Geomechanics for Energy and the Environment}, 28:\penalty0
  100255, 2021.

\bibitem[Khamitov et~al.(2022)Khamitov, Minh, and Zhao]{khamitov2022numerical}
Furkhat Khamitov, Nguyen~Hop Minh, and Yong Zhao.
\newblock Numerical investigation of sand production mechanisms in weak
  sandstone formations with various reservoir fluids.
\newblock \emph{International Journal of Rock Mechanics and Mining Sciences},
  154:\penalty0 105096, 2022.

\bibitem[Kloss et~al.(2012)Kloss, Goniva, Hager, Amberger, and
  Pirker]{kloss2012models}
Christoph Kloss, Christoph Goniva, Alice Hager, Stefan Amberger, and Stefan
  Pirker.
\newblock Models, algorithms and validation for opensource dem and cfd--dem.
\newblock \emph{Progress in Computational Fluid Dynamics, an International
  Journal}, 12\penalty0 (2-3):\penalty0 140--152, 2012.

\bibitem[Kooijman et~al.(1996)Kooijman, Van~den Hoek, De~Bree, Kenter, Zheng,
  and Khodaverdian]{kooijman1996horizontal}
AP~Kooijman, PJ~Van~den Hoek, Ph~De~Bree, CJ~Kenter, Z~Zheng, and
  M~Khodaverdian.
\newblock Horizontal wellbore stability and sand production in weakly
  consolidated sandstones.
\newblock In \emph{SPE Annual Technical Conference and Exhibition}. OnePetro,
  1996.

\bibitem[Kozhagulova et~al.(2021)Kozhagulova, Shabdirova, Minh, and
  Zhao]{kozhagulova2021integrated}
Ashirgul Kozhagulova, Ainash Shabdirova, Nguyen~Hop Minh, and Yong Zhao.
\newblock An integrated laboratory experiment of realistic diagenesis,
  perforation and sand production using a large artificial sandstone specimen.
\newblock \emph{Journal of Rock Mechanics and Geotechnical Engineering},
  13\penalty0 (1):\penalty0 154--166, 2021.

\bibitem[Lenormand et~al.(1988)Lenormand, Touboul, and
  Zarcone]{lenormand_touboul_zarcone_1988}
Roland Lenormand, Eric Touboul, and Cesar Zarcone.
\newblock Numerical models and experiments on immiscible displacements in
  porous media.
\newblock \emph{Journal of Fluid Mechanics}, 189:\penalty0 165–187, 1988.
\newblock \doi{10.1017/S0022112088000953}.

\bibitem[Li et~al.(2019)Li, Wu, Ning, Hu, Liu, Dong, and Lu]{li2019sand}
Yan-long Li, Neng-you Wu, Fu-long Ning, Gao-wei Hu, Chang-ling Liu, Chang-yin
  Dong, and Jing-an Lu.
\newblock A sand-production control system for gas production from clayey silt
  hydrate reservoirs.
\newblock \emph{China Geology}, 2\penalty0 (2):\penalty0 121--132, 2019.

\bibitem[Li et~al.(2005)Li, Xu, and Thornton]{li2005comparison}
Yanjie Li, Yong Xu, and Colin Thornton.
\newblock A comparison of discrete element simulations and experiments for
  ‘sandpiles’ composed of spherical particles.
\newblock \emph{Powder Technology}, 160\penalty0 (3):\penalty0 219--228, 2005.

\bibitem[Li et~al.(2023)Li, Espinoza, and Balhoff]{li2023simulation}
Zihao Li, D~Nicolas Espinoza, and Matthew~T Balhoff.
\newblock Simulation of polymer injection in granular media: implications of
  fluid-driven fractures, water quality, and undissolved polymers on polymer
  injectivity.
\newblock \emph{SPE Journal}, 28\penalty0 (01):\penalty0 289--300, 2023.

\bibitem[Ma et~al.(2021)Ma, Feng, Lin, Deng, Li, and Liu]{ma2021cfd}
Chengyun Ma, Yongcun Feng, Hai Lin, Jingen Deng, Xiaorong Li, and Fangrao Liu.
\newblock Cfd-dem investigation of blocking mechanism in pre-packed gravel
  screen.
\newblock \emph{Engineering Analysis with Boundary Elements}, 132:\penalty0
  416--426, 2021.

\bibitem[Mandal(2015)]{mandal2015chemical}
Ajay Mandal.
\newblock Chemical flood enhanced oil recovery: a review.
\newblock \emph{International Journal of Oil, Gas and Coal Technology},
  9\penalty0 (3):\penalty0 241--264, 2015.

\bibitem[Mindlin(1949)]{mindlin1949compliance}
Raymond~David Mindlin.
\newblock Compliance of elastic bodies in contact.
\newblock 1949.

\bibitem[Omirbekov et~al.(2023)Omirbekov, Colombano, Alamooti, Batikh,
  Cochennec, Amanbek, Ahmadi-Senichault, and
  Davarzani]{omirbekov2023experimental}
Sagyn Omirbekov, St{\'e}fan Colombano, Amir Alamooti, Ali Batikh, Maxime
  Cochennec, Yerlan Amanbek, Azita Ahmadi-Senichault, and Hossein Davarzani.
\newblock Experimental study of dnapl displacement by a new densified polymer
  solution and upscaling problems of aqueous polymer flow in porous media.
\newblock \emph{Journal of Contaminant Hydrology}, 252:\penalty0 104120, 2023.

\bibitem[Papamichos et~al.(2001)Papamichos, Vardoulakis, Tronvoll, and
  Skjaerstein]{papamichos2001volumetric}
Euripides Papamichos, I~Vardoulakis, J~Tronvoll, and A~Skjaerstein.
\newblock Volumetric sand production model and experiment.
\newblock \emph{International journal for numerical and analytical methods in
  geomechanics}, 25\penalty0 (8):\penalty0 789--808, 2001.

\bibitem[Qiu et~al.(2022)Qiu, Guo, Jin, Zhang, Si, and Guo]{qiu2022calibration}
Yiqing Qiu, Zhijun Guo, Xin Jin, Pangang Zhang, Shengjie Si, and Fugui Guo.
\newblock Calibration and verification test of cinnamon soil simulation
  parameters based on discrete element method.
\newblock \emph{Agriculture}, 12\penalty0 (8):\penalty0 1082, 2022.

\bibitem[Rahmati et~al.(2013)Rahmati, Jafarpour, Azadbakht, Nouri, Vaziri,
  Chan, and Xiao]{rahmati2013review}
Hossein Rahmati, Mahshid Jafarpour, Saman Azadbakht, Alireza Nouri, Hans
  Vaziri, Dave Chan, and Yuxing Xiao.
\newblock Review of sand production prediction models.
\newblock \emph{Journal of Petroleum Engineering}, 2013, 2013.

\bibitem[Rakhimzhanova et~al.(2021)Rakhimzhanova, Thornton, Amanbek, and
  Zhao]{rakhimzhanova2021numerical}
Aigerim Rakhimzhanova, Colin Thornton, Yerlan Amanbek, and Yong Zhao.
\newblock Numerical simulations of cone penetration in cemented sandstone.
\newblock In \emph{EPJ Web of Conferences}, volume 249, page 14010. EDP
  Sciences, 2021.

\bibitem[Rakhimzhanova et~al.(2022)Rakhimzhanova, Thornton, Amanbek, and
  Zhao]{rakhimzhanova2022numerical}
Aigerim Rakhimzhanova, Colin Thornton, Yerlan Amanbek, and Yong Zhao.
\newblock Numerical simulations of sand production in oil wells using the
  cfd-dem-ibm approach.
\newblock \emph{Journal of Petroleum Science and Engineering}, 208:\penalty0
  109529, 2022.

\bibitem[Rakhimzhanova et~al.(2019)Rakhimzhanova, Thornton, Minh, Fok, and
  Zhao]{rakhimzhanova2019numerical}
Aigerim~K Rakhimzhanova, Colin Thornton, Nguyen~Hop Minh, Sai~Cheong Fok, and
  Yong Zhao.
\newblock Numerical simulations of triaxial compression tests of cemented
  sandstone.
\newblock \emph{Computers and Geotechnics}, 113:\penalty0 103068, 2019.

\bibitem[Renaud et~al.(2004)Renaud, Mauret, and Chhabra]{renaud2004power}
Maurice Renaud, Evelyne Mauret, and Rajendra~P Chhabra.
\newblock Power-law fluid flow over a sphere: Average shear rate and drag
  coefficient.
\newblock \emph{The Canadian Journal of Chemical Engineering}, 82\penalty0
  (5):\penalty0 1066--1070, 2004.

\bibitem[Shabdirova et~al.(2020)Shabdirova, Khamitov, Kozhagulova, Amanbek,
  Minh, and Zhao]{shabdirova2020experimental}
AD~Shabdirova, Furkhat Khamitov, AA~Kozhagulova, Yerlan Amanbek, NH~Minh, and
  Yong Zhao.
\newblock Experimental and numerical investigation of the plastic zone
  permeability.
\newblock In \emph{54th US Rock Mechanics/Geomechanics Symposium}. OnePetro,
  2020.

\bibitem[Shahzad et~al.(2018)Shahzad, Aeken, Mottaghi, Kamyab, and
  Kuhn]{shahzad2018aggregation}
Khurram Shahzad, Wouter~Van Aeken, Milad Mottaghi, Vahid~Kazemi Kamyab, and
  Simon Kuhn.
\newblock Aggregation and clogging phenomena of rigid microparticles in
  microfluidics: Comparison of a discrete element method (dem) and cfd--dem
  coupling method.
\newblock \emph{Microfluidics and nanofluidics}, 22:\penalty0 1--17, 2018.

\bibitem[Terzaghi(1936)]{terzaghi1936stress}
Karl Terzaghi.
\newblock Stress distribution in dry and in saturated sand above a yielding
  trap-door.
\newblock 1936.

\bibitem[Tsuji et~al.(1993)Tsuji, Kawaguchi, and Tanaka]{tsuji1993discrete}
Yutaka Tsuji, Toshihiro Kawaguchi, and Toshitsugu Tanaka.
\newblock Discrete particle simulation of two-dimensional fluidized bed.
\newblock \emph{Powder technology}, 77\penalty0 (1):\penalty0 79--87, 1993.

\bibitem[Veeken et~al.(1991)Veeken, Davies, Kenter, and
  Kooijman]{veeken1991sand}
CAM Veeken, DR~Davies, CJ~Kenter, and AP~Kooijman.
\newblock Sand production prediction review: developing an integrated approach.
\newblock In \emph{SPE annual technical conference and exhibition}. OnePetro,
  1991.

\bibitem[Wang and Wu(2001)]{wang2001borehole}
Y~Wang and B~Wu.
\newblock Borehole collapse and sand production evaluation: Experimental
  testing, analytical solutions and field implications.
\newblock In \emph{DC Rocks 2001, The 38th US Symposium on Rock Mechanics
  (USRMS)}. OnePetro, 2001.

\bibitem[Wang et~al.(2022)Wang, Cheng, Yang, Tao, and Li]{wang2022microscopic}
Yin Wang, Kuang Cheng, Yefeng Yang, Yichen Tao, and Yewei Li.
\newblock Microscopic mechanical analysis of sand production using a new
  arbitrary resolved-unresolved cfd-dem model.
\newblock \emph{International Journal of Multiphase Flow}, 149:\penalty0
  103979, 2022.

\bibitem[Willson et~al.(2002)Willson, Moschovidis, Cameron, and
  Palmer]{willson2002new}
SM~Willson, ZA~Moschovidis, JR~Cameron, and ID~Palmer.
\newblock New model for predicting the rate of sand production.
\newblock In \emph{SPE/ISRM rock mechanics conference}. OnePetro, 2002.

\bibitem[Zhang et~al.(2019)Zhang, Wu, and Sharma]{zhang2019proppant}
Min Zhang, Chu-Hsiang Wu, and Mukul Sharma.
\newblock Proppant placement in perforation clusters in deviated wellbores.
\newblock In \emph{SPE/AAPG/SEG Unconventional Resources Technology
  Conference}. OnePetro, 2019.

\bibitem[Zhao and Shan(2013)]{zhao2013coupled}
Jidong Zhao and Tong Shan.
\newblock Coupled cfd--dem simulation of fluid--particle interaction in
  geomechanics.
\newblock \emph{Powder technology}, 239:\penalty0 248--258, 2013.

\bibitem[Zhou et~al.(2022)Zhou, Zhang, Hu, Li, Tang, Mao, and
  Wang]{zhou2022calibration}
Jiacheng Zhou, Libin Zhang, Chao Hu, Zhihang Li, Junjie Tang, Kuanmin Mao, and
  Xiaoyu Wang.
\newblock Calibration of wet sand and gravel particles based on jkr contact
  model.
\newblock \emph{Powder Technology}, 397:\penalty0 117005, 2022.

\bibitem[Zhou et~al.(2011)Zhou, Yu, and Choi]{zhou2011numerical}
ZY~Zhou, AB~Yu, and SK~Choi.
\newblock Numerical simulation of the liquid-induced erosion in a weakly bonded
  sand assembly.
\newblock \emph{Powder technology}, 211\penalty0 (2-3):\penalty0 237--249,
  2011.

\bibitem[Zhu et~al.(2007)Zhu, Zhou, Yang, and Yu]{zhu2007discrete}
HP~Zhu, ZY~Zhou, RY~Yang, and AB~Yu.
\newblock Discrete particle simulation of particulate systems: theoretical
  developments.
\newblock \emph{Chemical Engineering Science}, 62\penalty0 (13):\penalty0
  3378--3396, 2007.

\end{thebibliography}

\end{document}